\documentclass[useAMS,usenatbib]{mnras}

\usepackage{relsize}
\usepackage{josty}
\usepackage{amsmath}
\usepackage{graphicx}
\usepackage{animate}
\usepackage{amssymb}
\usepackage{comment}
 
\def\tq{\tau_q}
\def\tt{\tau_t}

\title[Quenching Timescales in IllustrisTNG] 
{Quenching Timescales in the IllustrisTNG Simulation}
\author[D. Walters et al.] 
    {\parbox{\textwidth}{Dan Walters,$^{1}$\thanks{danwalters@uvic.ca}
    Joanna Woo,$^{2}$
    and Sara L. Ellison,$^{1}$}
\vspace{0.4cm}\\
\parbox{\textwidth}{ 
$^{1}$Department of Physics \& Astronomy, University of Victoria, PO Box 1700 STN CSC, Victoria BC V8W 2Y2, Canada\\
$^{2}$Department of Physics, Simon Fraser University, 8888 University Drive, Burnaby BC V5A 1S6, Canada\\
}}

\pagerange{\pageref{firstpage}--\pageref{lastpage}} \pubyear{2021}

\hypersetup{draft}
\begin{document}
\label{firstpage}
\maketitle
 
\begin{abstract}
The timescales for galaxy quenching offer clues to its underlying physical drivers.  We investigate central galaxy quenching timescales in the IllustrisTNG 100-1 simulation, their evolution over time, and the pre-quenching properties of galaxies that predict their quenching timescales.  Defining quenching duration $\tq$ as the time between crossing sSFR thresholds, we find that $\sim40\%$ of galaxies quench rapidly with $\tq<$1 Gyr, but a substantial tail of galaxies can take up to 10 Gyr to quench.  Furthermore, 29\% of galaxies that left the star forming main sequence (SFMS) more than 2 Gyr ago never fully quench by $z=0$. While the median $\tq$ is {fairly} constant with epoch, the rate of galaxies leaving the SFMS increases steadily over cosmic time, with the rate of slow quenchers being dominant around $z\sim2$ to 0.7.  Compared to fast quenchers ($\tq<$1 Gyr), slow-quenching galaxies ($\tq>$1 Gyr) were more massive, had more massive black holes, had larger stellar radii and accreted gas with higher specific angular momentum (AM) prior to quenching. These properties evolve little by $z=0$, except for the accreting gas AM for fast quenchers, which reaches the same high AM as the gas in slow quenchers.  By $z=0$, slow quenchers also have residual star formation in extended gas rings. Using the expected relationship between stellar age gradient and $\tq$ for inside-out quenching we find agreement with MaNGA IFU observations. Our results suggest the accreting gas AM and potential well depth determine the quenching timescale. 
\end{abstract}


\section{Introduction}
\label{secintroduction}

Observations of local galaxy properties have shown that galaxies are bimodal in many properties, such as colour (e.g. \citealp{Strateva2001a,Blanton2003,Baldry2004}), stellar age (e.g. \citealp{Kauffmann2003b,Gallazzi2008,Peng2015a}), and morphology (e.g. \citealp{Fisher2011,Bottrell2019,Luo2020}), but most importantly, star formation rate (SFR) (e.g. \citealp{Mcgee2011,Wetzel2012a}). Galaxies can be broadly classified into two categories: star forming (SF) and quiescent (or passive/quenched, Q). Between SF and Q galaxies lie Green Valley (GV) galaxies \citep{Wyder2007}, although growing evidence against a true bimodality in SFR suggests that GV and Q galaxies may form a single extended population (e.g. \citealp{Feldmann2017,Terrazas2017,Eales2018a,Eales2018}). 
Galaxies form and grow along an SF main sequence (SFMS), with more massive SF galaxies having higher SFR (\citealp{Noeske2007a}). This evolution is stochastic, typically including multiple movements on and off the SFMS (e.g., \citealp{Tacchella2016a,Orr2017}). Regardless, galaxies generally seem to evolve along the SFMS, until at some point they are forced off the SFMS, rarely to return (the ``grow \& quench" framework, e.g. \citealp{Tacchella2021}). The process of transitioning off the SFMS is known as quenching, and it is a central focus in the study of galaxy evolution.

The causes of quenching are still poorly understood.  Different populations of galaxies exhibit different correlations between various galaxy properties and quiescence, which are thought to point to different quenching mechanisms \citep{Baldry2004,Bamford2009,Woo2015,Bluck2016}. In particular, galaxies are often divided into centrals (those that are isolated, or lie at the centre of a group or cluster's potential) and satellites (those that orbit centrals). Satellite galaxies are thought to quench via a variety of environmental processes which probably do not affect centrals, such as ram pressure stripping (e.g. \citealp{Abadi1999}), strangulation (e.g. \citealp{VanDenBosch2008}), or harassment (e.g. \citealp{Moore1996}). 

In this paper we focus on the quenching of central galaxies for which several possible mechanisms have been proposed. Most mechanisms focus on depleting a galaxy's supply of gas, either by preventing gas inflow or by ejecting it. In the most massive galaxies, accreting gas can become shock heated, preventing cooling and collapse (e.g. \citealp{Dekel2006}). Feedback from star formation and supernovae can expel gas in low-mass galaxies (e.g. \citealp{Dekel1986,Dekel2003}), but is likely not strong enough alone to induce galaxy-wide quenching in massive galaxies (e.g. \citealp{Su2019}). Feedback from accreting supermassive black holes (BHs), known as active galactic nuclei (AGN), can eject gas from a galaxy's core while also heating it and preventing infall (e.g. \citealp{Croton2006,Zinger2020}), although the strongest effects may be limited to the central few kpc \citep{Ellison2021}. Morphological quenching is  another possibility, whereby a dense galaxy core creates a steep potential, which in turn smooths the density of gaseous discs \citep{Martig2009}. This prevents the clumping of the gas necessary for star formation to occur. Galaxy mergers/interactions may play a role by triggering bursts of star formation or AGN activity (e.g. \citealp{Ellison2011,Knapen2015, Ellison2019, Hani2020}), although \cite{Quai2021} found that mergers in TNG do not lead to significant populations of quenched galaxies. 
{{Another newly proposed quenching mechanism is angular momentum quenching \citep{Peng2020,Renzini2020}. In this scenario, based on established scaling relations, gas accreting to high mass galaxies (particularly later in the universe) is prevented from cooling and clumping due to excessive angular momentum.}}
It is safe to say, out of the many proposed quenching mechanisms, AGN feedback remains the most popular, with many simulations relying on some form of AGN feedback in order to reproduce the observed quiescent population (e.g. \citealp{Weinberger2018,Schaye2015}).  Yet it has become increasingly accepted that several mechanisms may play a role (e.g. \citealp{Woo2015,Trussler2020,Tacchella2021}). 

The timeline of the quenching process can offer clues to the physical mechanisms at play.  For example, purely ejective mechanisms should act on shorter timescales comparable to the dynamical time, whereas preventative feedback should lead to longer timescales comparable to the depletion timescale (e.g., \citealp{Schawinski2014,Hahn2017}). Both the duration of quenching and its dependence on epoch remain largely unconstrained. Since we cannot directly observe the evolution of galaxies, determining quenching timescales accurately is inherently difficult. 

It has been argued, based on the dearth of galaxies in the GV, that quenching might be a fast process (relative to the age of the universe) (e.g. \citealp{Hahn2017}).  However other interpretations of the apparent GV are possible. For example, \cite{Eales2018a, Eales2018} argue that the apparent dearth of observed low-z GV galaxies (in terms of colour) is due to an overabundance of red galaxies, since galaxies in a wide range of low specific SFR (sSFR) all map to red colours. They show that the sSFR distribution is smooth rather than bimodal, weakening the argument that galaxies must pass quickly through the GV. They further argue that many red galaxies in fact continue to form stars at low rates. Likewise, \cite{Terrazas2017} found that low-z galaxies show a smooth distribution in $M_{BH}/M_*$, and proposed that the GV is populated with both transitioning and steady-state galaxies. In a similar vein, \cite{McDonald2016} examined a sample of brightest cluster galaxies (BCGs), which harbour the most massive BHs \citep{McConnell2013}. \cite{McDonald2016} found that BCGs in the local universe continue to form stars at low rates (below the widely used sSFR cutoff of $10^{-11}$ yr$^{-1}$ for SF galaxies) and their sSFRs have been gradually decreasing over very long timescales ($\sim$9 Gyr). \cite{Schawinski2014a} studied the morphologies of local galaxies, and argued through a study of their SEDs that quiescent galaxies with different morphologies must transition through the optical GV on very different timescales. Indeed, the appearance of a shallow GV is partly due to the common practice of combining SFR measurements with SFR upper limits in the same histograms. \cite{Feldmann2017} showed that the sSFR distribution in low- and high-z observations, as well as simulations, can be better fit by zero-inflated negative binomial distributions, rather than the commonly used overlapping log-normal distributions.

The duration of quenching is expected to depend on both epoch and mass, but determining the direction of these correlations is not at all trivial.  For example, if we consider quiescent galaxies at all redshifts and ask how long it took these galaxies to quench, we will find that the quenching timescale is shorter for quiescent galaxies observed at earlier times simply due to the limit of the age of the universe at the time of observation.  Observations of stellar populations of quiescent galaxies \citep{Goncalves2012,Tacchella2021} and post-starburst galaxies \citep{Wild2016,Suess2021} seem to corroborate this idea.
If however we take quiescent galaxies at $z=0$ and measure quenching timescales as a function of when they {\it began} to quench, those that began to quench earlier might on average have longer quenching timescales simply because they have more time to quench than a galaxy that began to quench recently.  Therefore the epoch-dependence of quenching timescales depends on whether the ``epoch" refers to the beginning or end of quenching.  To our knowledge, this subtlety has not been sufficiently appreciated in studies of quenching timescales.

Related to the question of quenching timescale as a function of epoch is the question of mass dependence.  It has been established that more massive galaxies appear to finish quenching earlier since the red sequence seems to be built from the top down \citep{Cowie1996,Bell2004,Faber2007,Tacchella2021}, a phenomenon called ``downsizing".  Given that more massive galaxies quench earlier, and that earlier (completed) quenching is more rapid, one might naively expect that more massive galaxies quench more rapidly.  However the mass-dependence of the quenching timescale is not at all straightforward.  Some simulation studies find that more massive galaxies quench more quickly \citep{Nelson2018a, Wright2018,Dave2019}, while some observational studies seem to indicate the opposite \citep{Peng2015a,Trussler2020}.  Still others find little to no mass correlation at all \citep{Tacchella2021}. 

As we show in this study, at least some of the discrepancies are likely due to differing definitions of quenching timescales.  For example, \cite{Nelson2018a} defined quenching in the IllustrisTNG simulation using $g-r$ colour instead of sSFR, the latter of which is more directly related to quenching.  As we show here, it turns out that measuring timescales by sSFR as opposed to by colour actually reverses the direction of mass-dependent quenching, particularly with respect to the dependence of quenching duration on $\Ms$.

In this paper, we study quenching timescales in the IllustrisTNG simulations (hereafter TNG) \citep{Nelson2018,Pillepich2018,Springel2018,Naiman2018,Marinacci2018,Nelson2019} as a function of epoch, mass and other galaxy properties. 
TNG is a set of magneto-hydrodynamical simulations covering large cosmological volumes with sub-kpc resolution. By tracking individual galaxy evolution in TNG, we aim to quantify quenching duration timescales and quenching rates over cosmic time.  Our main goal is to determine the galaxy properties that predict quenching duration, and the evolution of these properties over time. To complement our simulation study, we also use measurements of stellar age gradients \citep{Woo2019} from the Mapping Nearby Galaxies at the Apache Point Observatory (MaNGA) survey \citep{Bundy2015} of local quiescent galaxies to validate our TNG results.

This paper is structured as follows. In Section \ref{secmethods} we discuss our methods. In Section \ref{secdist}, we present the overall distribution of quenching durations in TNG. In Section \ref{secepoch} we investigate the evolution of quenching rates over cosmic time. We then examine the relationship between quenching duration and the properties of simulated galaxies immediately prior to leaving the SFMS in Section \ref{seclsf}. Next, in Section \ref{secz0} we explore galaxy properties at $z=0$ based on quenching duration. Finally, we discuss the implications of our results, compare our results with observations, and address tensions with other studies in Section \ref{secdiscussion}. We conclude in Section \ref{secsummary}.

\section{Methods}
\label{secmethods}

\subsection{Simulation}
\label{simmethod}

IllustrisTNG is a set of large-box cosmological gravo-magnetohydrodynamical simulations run from $z=127$ to $z=0$ \citep{Nelson2019}. TNG uses a $\Lambda$CDM cosmology with inital conditions and parameters consistent with Planck 2015 \citep{Ade2016}. Subgrid treatments include stochastic star formation, stellar evolution and feedback, and supermassive BH seeding, growth, and feedback. TNG is calibrated to reproduce several observed relations, notably the cosmic SFR density, galaxy stellar mass function, stellar size-stellar mass relation, and BH-stellar mass relation (for quiescent galaxies) \citep{Pillepich2018}. Subhaloes and halos are identified using the SUBFIND and friends-of-friends (FOF) algorithms respectively \citep{Springel2001}.

Notably, and in contrast with other simulations, TNG simulates AGN feedback with a combination of thermal (quasar mode) and isotropic kinetic (radio mode) feedback \citep{Weinberger2017}. AGNs in TNG are simulated by feedback from BH particles, which are essentially constrained to lie at the centres of subhaloes. BH particles accrete gas at the Bondi accretion rate up to the Eddington limit. Feedback from AGNs operates in either a thermal (quasar) mode or kinetic (radio) mode depending on accretion rate and $\MBH$ \citep{Weinberger2017}. At high accretion rates, the thermal feedback mode injects thermal energy into the surrounding gas. At low accretion rates and high $\MBH$, the kinetic feedback mode provides momentum kicks to the surrounding gas in random directions. The onset of quenching in TNG is strongly linked to kinetic mode feedback, which typically begins at $\MBH \sim 10^{8.2} \Msun$ \citep{Terrazas2020}. Kinetic mode feedback both ejects gas from the subhalo while also preventing infall \citep{Zinger2020}.

TNG's success in reproducing observed quenched fractions \citep{Donnari2020a,Donnari2020b} and the locus and shape of the SFMS \citep{Donnari2019} across stellar mass and redshift makes it particularly well suited to studying quenching phenomena. TNG is also notable in producing the observed split between SF and Q populations in the $\sigone$-$\Ms$ and $\MBH$-$\Ms$ planes \citep{Walters2021}, although BHs are quantitatively too massive in TNG's high-mass SF subhaloes \citep{Terrazas2016,Terrazas2020}. Finally, isolated TNG SF galaxies have similar radial sSFR, age, and metallicity profiles as observations \citep{Walters2021}. At least in TNG50, central galaxies quench inside-out \cite{Nelson2021}, in agreement with recent IFU observations \citep{Ellison2018,Lin2019,Bluck2020}.

There are three sets of TNG simulations, with comoving box sizes of 50$^3$ cMpc${^3}$, 100$^3$ cMpc${^3}$, and 300$^3$ cMpc${^3}$. In this study we use TNG100-1 which provides a good compromise between box size (100$^3$ cMpc${^3}$) and resolution (gravitational softening length of 0.74 ckpc). We note that TNG100-1 lacks rare high-mass objects (log $\Mhalo/\Msun > 14.5$) and is therefore not mass-complete \citep{Pillepich2018a}, although we do not expect this to significantly impact our results. We also used the lower resolution TNG100-2 to check for convergence of our results (not shown).

For our study, we select central subhaloes with $\Ms > 10^9 \Msun$ at $z=0$ (12227 total). We also require that our subhaloes are centrals in the snapshot immediately prior to quenching (i.e. they are not splashback galaxies that quenched as satellites).  To avoid issues with rapid quenching and rejuvenation at very early cosmic times when most galaxies are not well resolved, we do not include times prior to $z=3.5$ in our analysis ($\sim$ 12 Gyr ago). We exclude subhaloes that never appear on the SFMS since that time (only 2 subhaloes). Except when stated otherwise, global values (e.g. $\Ms$) include all SUBFIND bound particles. Values for central stellar surface density $\sigtwo$, stellar age gradients, and infalling gas properties are calculated following \cite{Walters2021}. Unless stated otherwise, subhalo histories are found by tracing back in time along the main progenitor branch. We note that we do not consider subhalo mergers in our analysis. \cite{Quai2021} showed that in the TNG model, mergers do not significantly contribute to quenching (\citealp{RodriguezMontero2019} found a similar result in the SIMBA simulations).

\subsection{Quenching Duration}
\label{quenchmethod}

We use a subhalo's sSFR to define its classification at a given epoch as SF, transitioning (GV - although strictly speaking our cuts are based on sSFR, not colour), or Q. \Fig{ssfr_anim} shows an animation of central subhaloes in the log sSFR-log $\Ms$ plane for the entire simulation. We define the locus of the SFMS (sSFR$_{\rm SFMS}$) as the median sSFR for subhaloes with $\Ms < 10^{10} \Msun$ (black dashed line). We define SF subhaloes as those with log sSFR $>$ log sSFR$_{\rm SFMS}$ - 0.7. We define Q subhaloes as those with log sSFR $<$ log sSFR$_{\rm SFMS}$ - 2.7, and include subhaloes with unresolved sSFR (sSFR = 0 in the simulation) in the Q category. Subhaloes between these two bounds are labelled transitioning (green shaded area in \Fig{ssfr_anim}). Our choice of 2.0 dex for the GV width is somewhat arbitrary. It ensures that we always have some resolved Q subhaloes below the lower bound (i.e. we are not up against a resolution limit at the lower end). Further, by choosing a fairly wide GV, we hope to gain the most possible information about variability in quenching durations. Our sample has 10189 SF, 1083 GV, and 955 Q  subhaloes at $z=0$. 

In \Fig{examples} we show some example trajectories of sSFR through time for individual subhaloes. It can be seen that these evolutionary tracks tend to be stochastic, but for the most part subhaloes evolve along the SFMS, until at some point they drop into the GV and below. Rejuvenation from fully quenched back to the SFMS is rare in TNG ($\sim8\%$ rejuvenate at least once since $z=3.5$ using our definitions). We did try varying our SF/GV/Q bounds. Smaller GV widths (e.g. 1.0 dex) generally result in shorter quenching times, but correlations between quenching duration other galaxy properties are not substantively affected.

\begin{figure}
   \animategraphics[loop,autoplay,width=1.0\linewidth]{12.5}{figures/ssfr_anim/cutoff_snap_}{00}{99} 
  \caption{\small Distribution of TNG100-1 central subhaloes in log sSFR-log $\Ms$ over cosmic time. The top panel shows subhaloes with resolved sSFR while the bottom panel is a histogram of those with sSFR below the numerical resolution limit of the simulation (i.e., sSFR = 0). The black dashed line indicates the locus of the SFMS. The green shaded area indicates the GV, from 0.7 dex to 2.7 dex below the SFMS locus. SF subhaloes lie above the green shaded area, GV subhaloes are within the green area, and Q subhaloes are the combination of the scatter points below the green area plus subhaloes in the red histogram. We define the beginning of quenching as the last time a subhalo sits above the GV. We define the end of quenching as the first time a subhalo reaches the bottom of the GV. The animation embedded in this PDF file can be viewed in Adobe Reader.}
\label{ssfr_anim}
\end{figure}

\begin{figure}
\includegraphics[width=1.0\linewidth]{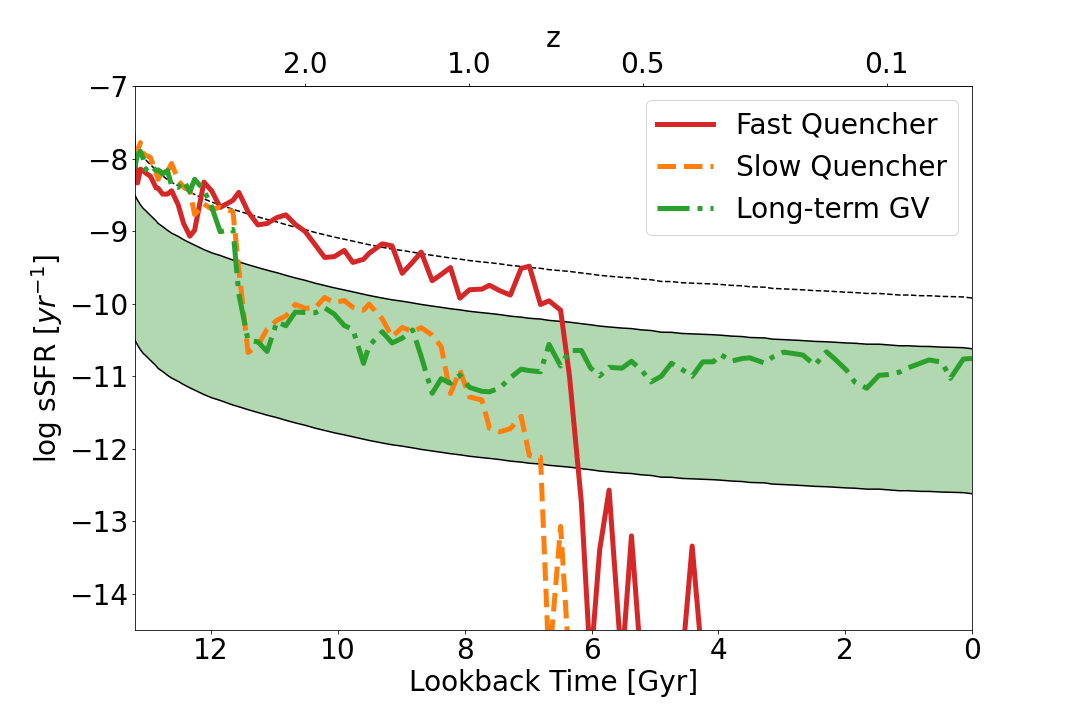}
\caption{\small sSFR history for three example subhaloes: a fast quencher (red, $\tq$ = 0.3 Gyr), a slow quencher (orange dashed, $\tq$ = 5.0 Gyr), and a long-term GV resident (green dash dotted, $\tt$ = 11.7 Gyr). The black dashed line shows the locus of the SFMS. The shaded green area shows the transition region (GV). Although histories are somewhat stochastic, almost all subhaloes in TNG form and evolve along the SFMS before leaving at a discrete point, never to return. Trajectories after leaving the SFMS vary greatly and not are well parameterized by, for example, exponentially decaying profiles.} 
\label{examples}
\end{figure}

We define the beginning of the quenching process as the last lookback time at which a subhalo sits in the SF region ($t_{\rm LSF}$) before its first quenching event. If a subhalo has multiple quenching events, we only consider the first one. All lookback times are from $z=0$. We define the end of quenching as the first time a subhalo reaches the bottom of the GV ($t_{\rm Q}$). We require that subhaloes be fully quenched for two consecutive snapshots to avoid momentary dips due to, for example, SUBFIND switching near mergers. The duration of quenching is $\tau_q = t_{\rm LSF} - t_{\rm Q}$. For those subhaloes that leave the SFMS, but never reach the lower quenching boundary and remain in the GV at $z=0$, we define the time in transition as $\tau_t = t_{\rm LSF}$. The time resolution is limited by the spacing of snapshots in TNG, which varies non-monotonically between 87 and 235 Myr after $z \sim 4$. Furthermore, we tag a subhalo as rejuvenated if it climbs back to the SFMS at any point after its first quenching event.

\section{Distribution of Quenching Durations}
\label{secdist}

\begin{figure}
\includegraphics[width=1.0\linewidth]{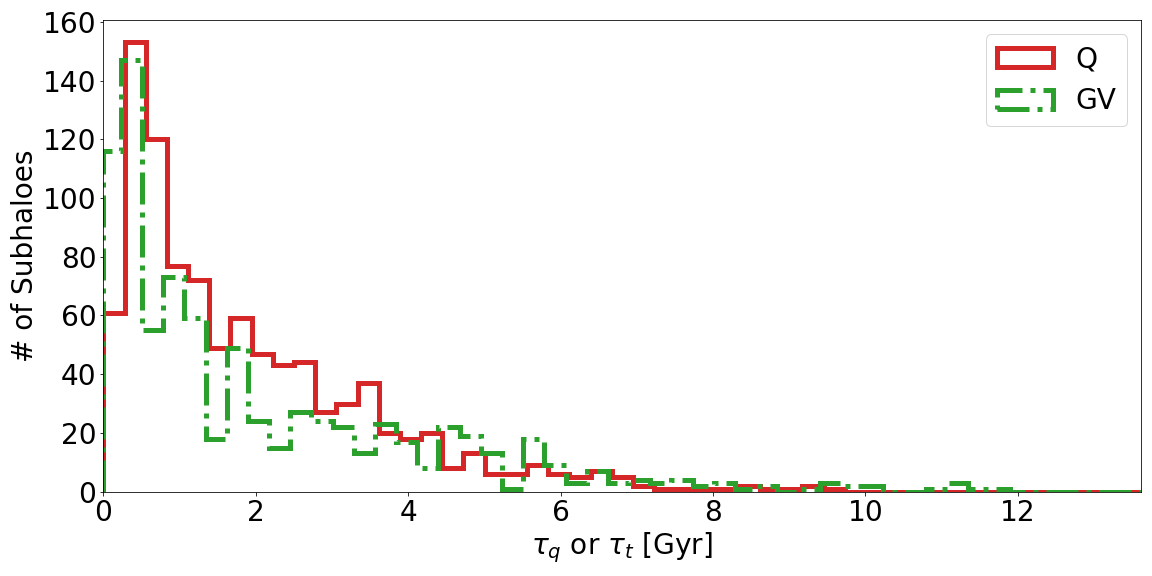}
\caption{\small Quenching duration $\tq$ for all TNG100-1 centrals ($\Ms > 10^9 \Msun$) that have ever quenched by $z=0$ (red, 955 total) and time since leaving the SFMS $\tt$ for all centrals that are in the GV at $z=0$ and never fully quenched (green dash dotted, 820 total). Many subhaloes quench quickly ($\lesssim$1 Gyr), but a significant number quench slowly, in agreement with Nelson et al. (2018). The GV is populated both with subhaloes in rapid transition to quiescence and with long-term 'residents' (or equivalently, very slow quenchers).} 
\label{tauq_dist}
\end{figure}

We begin by presenting the overall distributions of $\tq$ (the quenching duration for galaxies that have fully quenched at least once)  and $\tt$ (the time in transition for galaxies that have left the SFMS, never fully quenched, and are still in the GV by $z=0$) in TNG. The red line in \Fig{tauq_dist} shows $\tq$ for our sample. Many subhaloes ($\sim40\%$) quench quickly ($<$ 1 Gyr), but a significant tail of subhaloes with longer quenching times (up to $\sim$ 10 Gyr) exists, in agreement with the TNG results of \cite{Nelson2018a}.  38\% of quiescent central galaxies took more than 2 Gyr to quench (16\% with a narrower 1-dex GV).  The median $\tq$ is 1.3 Gyr (0.7 Gyr with a 1-dex GV). This is broadly consistent with the recent observational results of \cite{Tacchella2021} at $z>0.8$, who found $\tq\approx$ 0-5 Gyr with a median of 1.0 Gyr, although they use a different definition of $\tq$. We caution that exact $\tq$ values less than 1 Gyr are unreliable due to the snapshot spacing of TNG, however we can confidently state that these subhaloes quenched in $<$ 1 Gyr.

The green histogram in \Fig{tauq_dist} shows the transition time, $\tt$, for all centrals that are in the GV at $z=0$ and never fully quenched. The majority of subhaloes that have been in the GV for $\lesssim$ 1-2 Gyr are likely in the process of transitioning to quiescence. However, 29\% of transitioning galaxies left the SFMS more than 2 Gyr ago, never fully quenched but are still showing non-negligible SFR at $z=0$.  A few galaxies (8 total, $\sim$ 1\%) have been in the GV for over 10 Gyr. Galaxies in TNG can therefore clearly inhabit the GV for a very extended period of time, contrary to the idea of a quick transition from the SFMS to the passive regime.

The long-lived nature of GV galaxies is even more stark when we trace subhalo histories forward in time. Of those in transition at $z=2$ ($\sim$ 10 Gyr ago), 49\% remain in the GV at $z=0$ and do not fully quench.  If we define a narrower GV width of 1.0 dex, the number is smaller (25\%) but still significant. 

The exact percentages and timescales are of course dependent on the width of the transition range of sSFR. Nevertheless, we qualitatively conclude that in TNG the GV is populated both with subhaloes in a true transition to quiescence as well as with subhaloes `residing' in the GV.  These findings support the proposal of \cite{Terrazas2017} that many galaxies may live in a ``quasi-stable" state of ``partial quiescence." In summary TNG produces a broad range of quenching times, whereas many quench in under a Gyr the slowest quenchers never fully cease to form stars within the age of the universe.

\section{Dependence on Cosmic Epoch}
\label{secepoch}

\begin{figure*}
    \centering
    \includegraphics[width=1.0\linewidth]{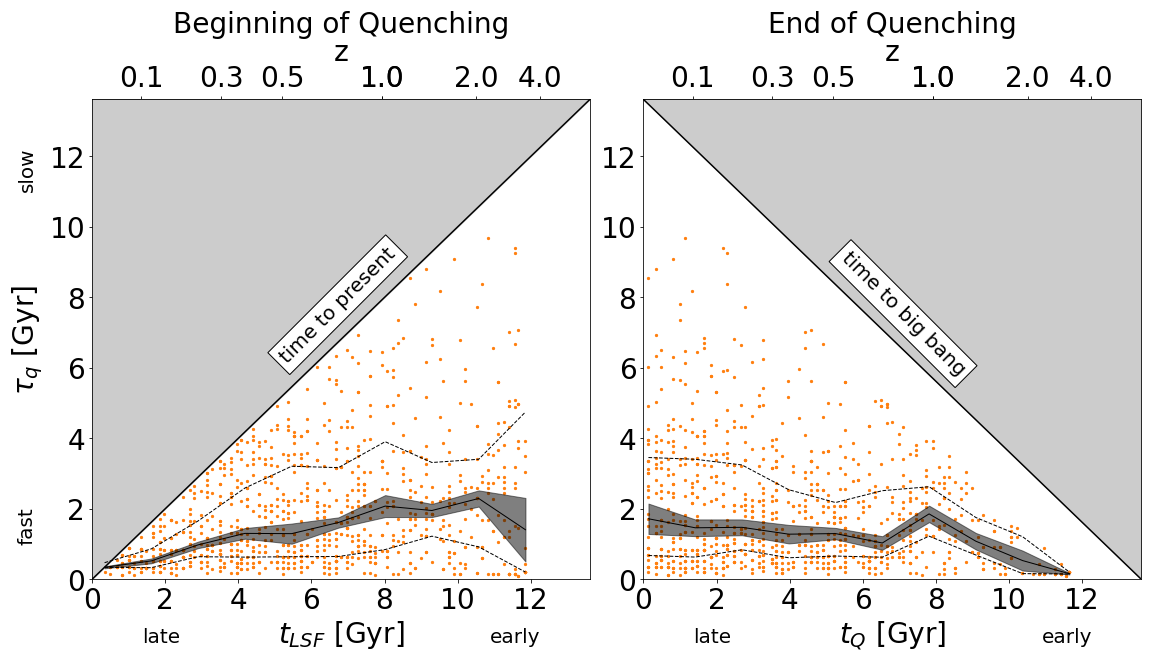}
    \caption{Quenching duration $\tq$ versus the lookback time at which each galaxy left the SFMS, $t_{\rm LSF}$ (left panel), or versus the lookback time at which each galaxy became fully quenched, $t_Q$ (right panel). The 1:1 lines indicate the maximum $\tq$ for a galaxy to quench by $z=0$ (left panel) or to have begun quenching after the birth of the universe (right panel). $\tq$ can never lie in the grey area above these lines. By definition, all $\tt$ values (i.e., the transition times for galaxies are still in the GV by $z=0$) lie on the diagonal line in the left panel. The black curve is a running median using a sliding bin $\pm0.5$ Gyr wide. The shaded error is the error on the median and the dashed lines are quartiles. When defining the epoch of quenching as the beginning of the quenching process (left panel), shorter $\tq$ values arise naturally at later times. Likewise, when defining the epoch of quenching as the end of process, shorter $\tq$ values arise at early times. {However, median $\tq$ values in TNG only vary weakly with cosmic time.} Unless stated otherwise, we use $t_{\rm LSF}$ (i.e. the left panel) as the epoch of quenching in this paper.}
    \label{tauq_vs_lookback}
\end{figure*}

Next, we investigate the dependence of quenching duration on cosmic epoch.  It is important to emphasize that here we refer to the epoch at which the galaxies begin to quench.  As argued in the Introduction, the maximum possible $\tq$ will decrease with the redshift of quenching ($z_q$) if $z_q$ refers to the time at which the galaxy completes its quenching, simply because of the smaller age of the universe at high $z_q$ (see for instance Fig. 13 of \citealp{Tacchella2021} which shows the upper physical bound to $\tq$ given their definition of $z_q$).  If instead the epoch $z_q$ refers to the time at which a galaxy leaves the SFMS, median $\tq$ values might be expected to increase with $z_q$ simply because galaxies that begin to quench earlier will have had more time to quench by the time they reach $z=0$.  We illustrate this idea in \Fig{tauq_vs_lookback} which shows $\tq$ for quiescent galaxies as a function of $t_{\rm LSF}$ (left panel; the lookback time of leaving the SFMS) and as a function of $t_Q$ (right panel; the lookback time of becoming fully quenched).  By definition, $\tt$ values for GV galaxies lie on the 1:1 diagonal line in the left panel. {For the quiescent galaxies the average $\tq$ does indeed decrease slightly at late times (left panel) or early times (right panel) due to the limited values possible, but is otherwise is surprisingly close to constant for most of the history of the universe. In other words, $\tq$ in TNG only weakly depends on the epoch at which quenching begins or ends, with any dependence plausibly explained by the constraints discussed above}.

\begin{figure}
\includegraphics[width=1.0\linewidth]{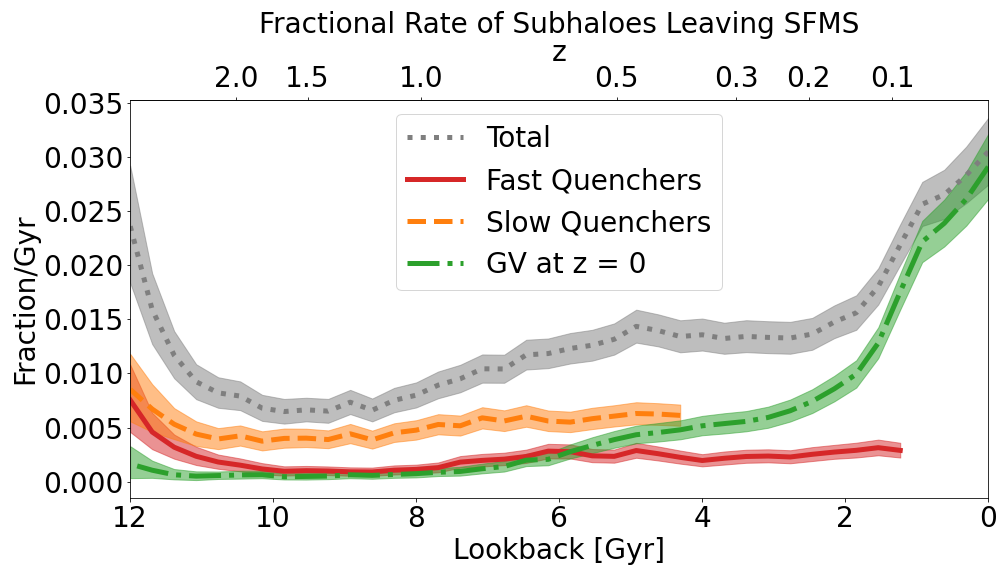}
\caption{\small Rate of subhaloes leaving the SFMS divided by the total number of subhaloes on the SFMS. The grey dotted line shows all subhaloes leaving the SFMS. All values are calculated using a sliding bin $\pm$1 Gyr wide. The width of the shaded area is the Poisson error in the bin. The overall quenching rate in TNG increases steadily over time after $z\sim2$. The other three lines show portions of the total quenching population based on their fate: fast quenchers (red solid), slow quenchers (orange dashed), and those that will reside in the GV until $z=0$ (green dash dotted). Slow quenchers leave the SFMS at about double the rate of fast quenchers from $\sim z=2$ to $\sim z=0.7$, at which point fast quenching becomes about as common. After $z\sim0.5$, the end of the simulation ($z=0$) prevents drawing meaningful conclusions about quenching rates.}
\label{epoch}
\end{figure}

In a similar vein, we next look at the rate at which certain types of quenchers leave the SFMS as a function of time.
\Fig{epoch} shows the fractional rate of subhaloes leaving the SFMS (either to fully quench, or to enter the GV and not return) over time.  We define the fractional rate as the number of subhaloes leaving the SFMS per unit time (over a sliding bin of $\pm$1 Gyr) divided by the median total number of subhaloes on the SFMS within the time bin above a mass of $10^9 \Msun$. The width of the shaded area is the Poisson error in the bin. The dotted grey curve shows the whole galaxy population of central subhaloes with $\Ms > 10^9 \Msun$. In TNG, after $z\sim2$, the total fractional quenching rate steadily increases (prior to $z\sim2$, the small number of total galaxies with $\Ms > 10^9 \Msun$ drives the fractional rate up). We note that the number of subhaloes on the SFMS is roughly constant after this point, and indeed the absolute quenching rate also steadily increases (not shown).  The bumps and plateaus are somewhat sensitive to our SFMS boundary, which itself evolves over time.  However the steady increase of the rate at which galaxies leave the SFMS with time is a robust prediction of TNG, regardless of our choice of SFMS width.  

The other three curves in \Fig{epoch} show components of the total, split by their fate: fast quenchers ($\tq < 1$ Gyr - red solid), slow quenchers (1 Gyr $< \tq <$ 4 Gyr - orange dashed), and those still in the GV at $z=0$ (green dash dotted). We note that these curves do not add up to the total fractional rate due to the range limits on $\tq$ for each class. For example, subhaloes that will take 3 Gyr to quench will not have time to do so if they leave the SFMS 2 Gyr ago. Thus we omit the very long quenchers ($\tq >$ 4 Gyr) from the plot, which limits the orange curve to extend only to a lookback time of $> 4$ Gyr.

At all epochs at $z<2$, slow quenching ($\tq >$ 1 Gyr) is more common than fast quenching ($\tq <$ 1 Gyr). From $z\sim2$ to $z\sim0.7$, \Fig{epoch} shows that slow quenchers leave the SFMS at about double the rate of fast quenchers. That is, slow quenchers are about twice as common as fast quenchers at intermediate redshifts. We also see the same trend if we use $t_Q$ instead of $t_{\rm LSF}$ (not shown). However the numbers of subhaloes quenching at these epochs are low.
After $z\sim0.5$, we do not know the fates of enough of the quenching subhaloes to make meaningful comparisons, at least for the slow quenchers.  However the rate of galaxies leaving the SFMS between 4 and 1 Gyr ago that are still in the GV today is well above the rate of the fast quenchers leaving the SFMS, indicating that slow quenching continues to dominate at late times.

\begin{figure}
    \centering
    \includegraphics[width=1.0\linewidth]{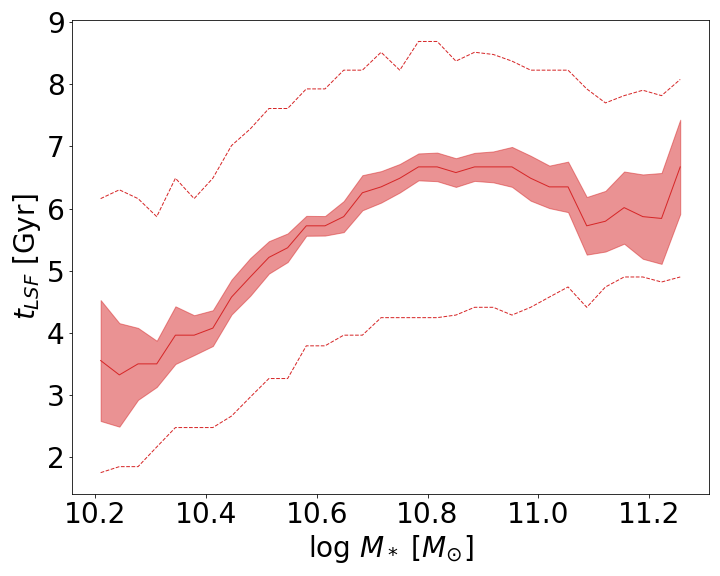}
    \caption{Lookback time that quenched galaxies were last on the SFMS ($t_{\rm LSF}$) versus stellar mass prior to leaving the SFMS. The solid line is a running median using a sliding box $\pm0.1$ dex wide, the shaded area is the error on the median, and the dashed lines are quartiles. The mass range is chosen to have at least 25 points in the bin. In TNG, more massive galaxies tend to leave the SFMS at earlier epochs.}
    \label{massepoch}
\end{figure}

How does quenching epoch depend on galaxy mass?  \fig{massepoch}  shows the median epoch of leaving the SFMS as a function of stellar mass.  This figure shows that the epoch at which galaxies leave the SFMS depends on mass such that more massive galaxies {\it begin} quenching earlier. In other words, not only do slower quenchers dominate at higher $z$, in TNG, more massive galaxies leave the SFMS earlier than less massive galaxies. {{We also find that more massive galaxies finish quenching earlier ($\sim$2 Gyr earlier on average, not shown for brevity), but the mass dependence is weaker.  These results are qualitatively consistent with observational studies of number density statistics that more massive galaxies {\it finish} quenching at earlier times \citep{Cowie1996,Bell2004,Faber2007, Tacchella2021}. Note however that observational studies are not directly comparable to our results because they are based on observed quiescent galaxies at different redshifts, which are upper limits to the epoch of quenching completion, and do not include galaxies that are in transition to quenching.}}

\section{Subhalo Properties Prior to Leaving the SFMS}
\label{seclsf}

Why do some TNG subhaloes quench quickly, some slowly, and some never fully quench at all? To answer this, we investigate which properties of the subhaloes prior to leaving the SFMS correlate with quenching duration.  

\begin{figure*}
    \centering
    \includegraphics[width=0.95\linewidth]{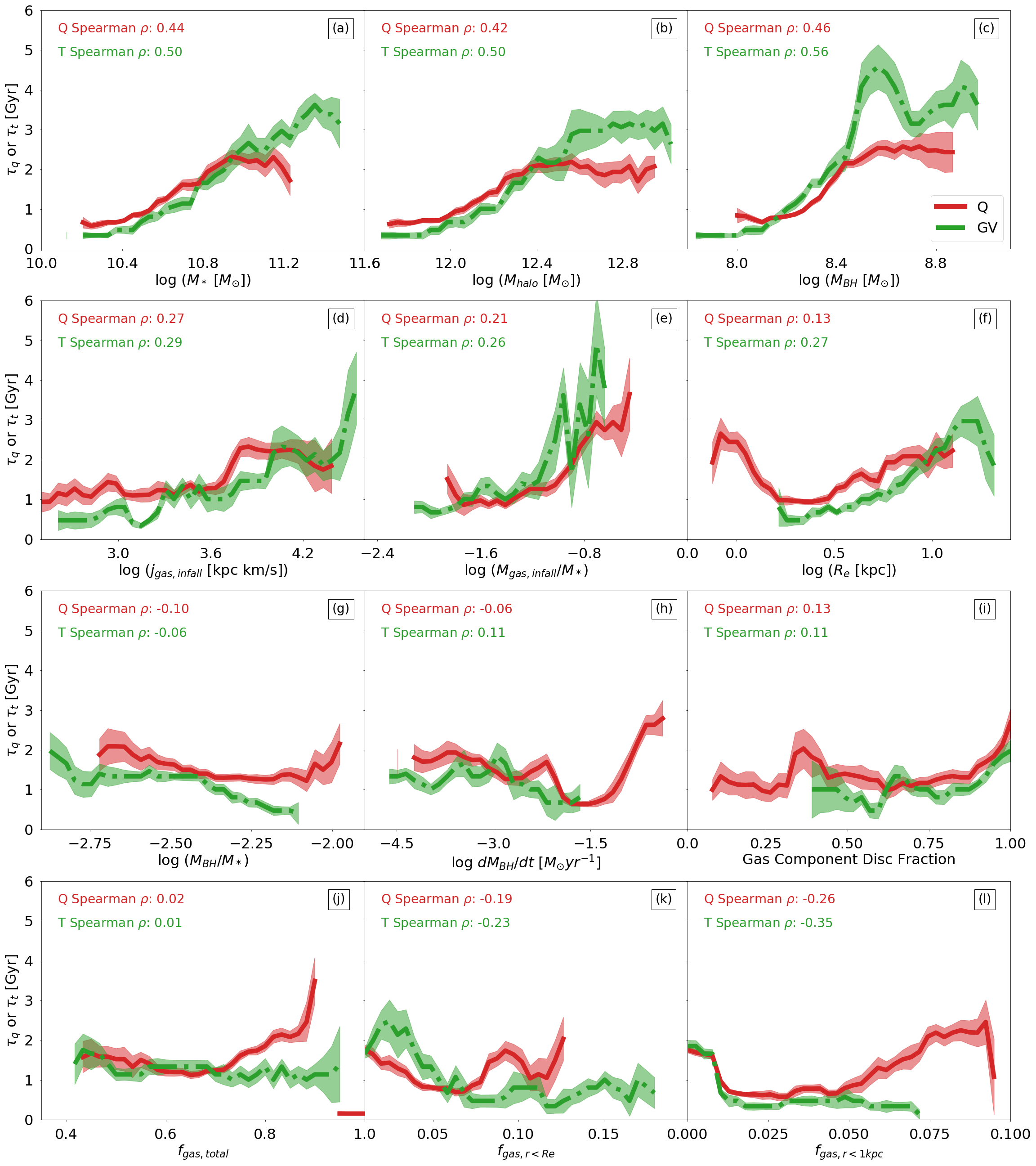}
    \caption{Correlation of $\tq$ (solid red) or $\tt$ (green dash dotted) with 12 subhalo properties taken immediately prior to leaving the SFMS: (a) stellar mass $\Ms$, (b) total halo mass $M_{\rm halo}$, (c) black hole mass $\MBH$, (d) infalling gas specific angular momentum ($j_{\rm gas,infall}$), (e) infalling gas mass ($M_{\rm gas,infall}$, normalized by $\Ms$), (f) stellar half-mass radius $R_e$, $\MBH$ normalized by $\Ms$, (g) the relative BH mass ($\MBH/\Ms$), (h) instantaneous BH accretion rate, (i) the disk fraction of the gas component, (j) the total SUBFIND gas fractions, (k) the gas fraction within Re, and (l) within 1 kpc. The solid/dash-dotted lines are the running medians and shaded areas are the error on the median, all calculated using a sliding bin. The width of the sliding bin is $\pm0.1$ dex (panels a-g), $\pm0.2$ dex (panel h), $\pm0.05$ (panels i and j), or $\pm0.01$ (panels k and l). The lines are not shown when there are fewer than 25 points in the bin. Red shows galaxies that have fully quenched at some point in their history, with y-axis values indicating quenching duration ($\tq$). Green shows galaxies that left the SFMS and are in the GV at $z=0$, with y-axis values being the length of time since leaving the SFMS ($\tt$). Low values of $\tt$ should be interpreted with caution as many of these galaxies began to quench at late times and could soon quench. Spearman correlation coefficients for each data set are shown in the top left of each panel. Quenching duration correlates strongest with the masses in panels a-c, and also moderately with infalling gas properties in panels d and e, and galaxy size in panel f. The fastest quenching in TNG is associated with lower $\Ms$, lower $M_{\rm halo}$, lower $\MBH$ and higher infalling gas angular momentum and mass.}
    \label{lsf_cont_all}
\end{figure*}

We begin by examining the correlation of $\tq$ and $\tt$ with 12 subhalo properties taken at the last snapshot prior to leaving the SFMS (\Fig{lsf_cont_all}). Each panel shows two running medians: the green dash dotted curve shows $\tt$ for subhaloes in transition at $z=0$ which have never fully quenched, and the solid red curve shows $\tq$ for subhaloes that have fully quenched at least once. Binning details for the running medians are in the caption.  Each panel is labelled with the Spearman correlation coefficient for each data set.

The first two panels (a and b) of \Fig{lsf_cont_all} show the stellar mass and the halo mass prior to leaving the SFMS. They show that in TNG, more massive galaxies living in more massive halos tend to take longer to quench. Despite the large scatter, TNG predicts a clear positive correlation between quenching duration and stellar mass. This result at first seems to be in conflict with the observations that more massive galaxies quench earlier \citep{Bell2004,Faber2007,Tacchella2021} and that earlier quenching is more rapid \citep{Goncalves2012,Suess2021}.  Furthermore, they are in conflict with the results of previous simulation work \citep{Nelson2018a,Wright2018,Dave2019}.  However, we will show in Section \ref{secdiscussion} that our results do in fact agree with observations, and that the discrepancies depend crucially on definitions of quenching timescale.  

Panel (c) of \Fig{lsf_cont_all} shows BH mass. Virtually all quiescent and long-term GV galaxies have BH masses higher than $\MBH \sim10^{8.0}\Msun$. At this mass, BHs tend to switch to kinetic feedback, which is the primary cause of quenching in TNG \citep{Terrazas2020}. Galaxies with higher BH masses prior to quenching tend to take longer to quench, which is consistent with the relation between $\Ms$ and $\tq$. This is contrary to the common view (e.g. \citealp{Nelson2018a}) that the most massive BHs drive the fastest quenching. 

These first three properties ($\Ms$, $M_{\rm halo}$, and $\MBH$) show the strongest correlation with $\tq$ and $\tt$ (all have Spearman $\rho >$ 0.4). We note that all three of these properties are also well known to correlate with each other. 

Panels (d) and (e) of \Fig{lsf_cont_all} examine the gas accreting to the subhalo from a large radius. We take all gas particles with position and inward radial velocity such that they will cross $R_{\rm vir}/2$ in the next 100 Myr, summing their specific angular momentum (panel d) and mass (panel e).  These panels show that the properties of the infalling gas are moderately correlated with $\tq$ and $\tt$ (Spearman $\rho >$ 0.2).  Galaxies accreting less gas, or gas with lower specific angular momentum, tend to quench quickly.

\begin{figure*}
\includegraphics[width=1.0\linewidth]{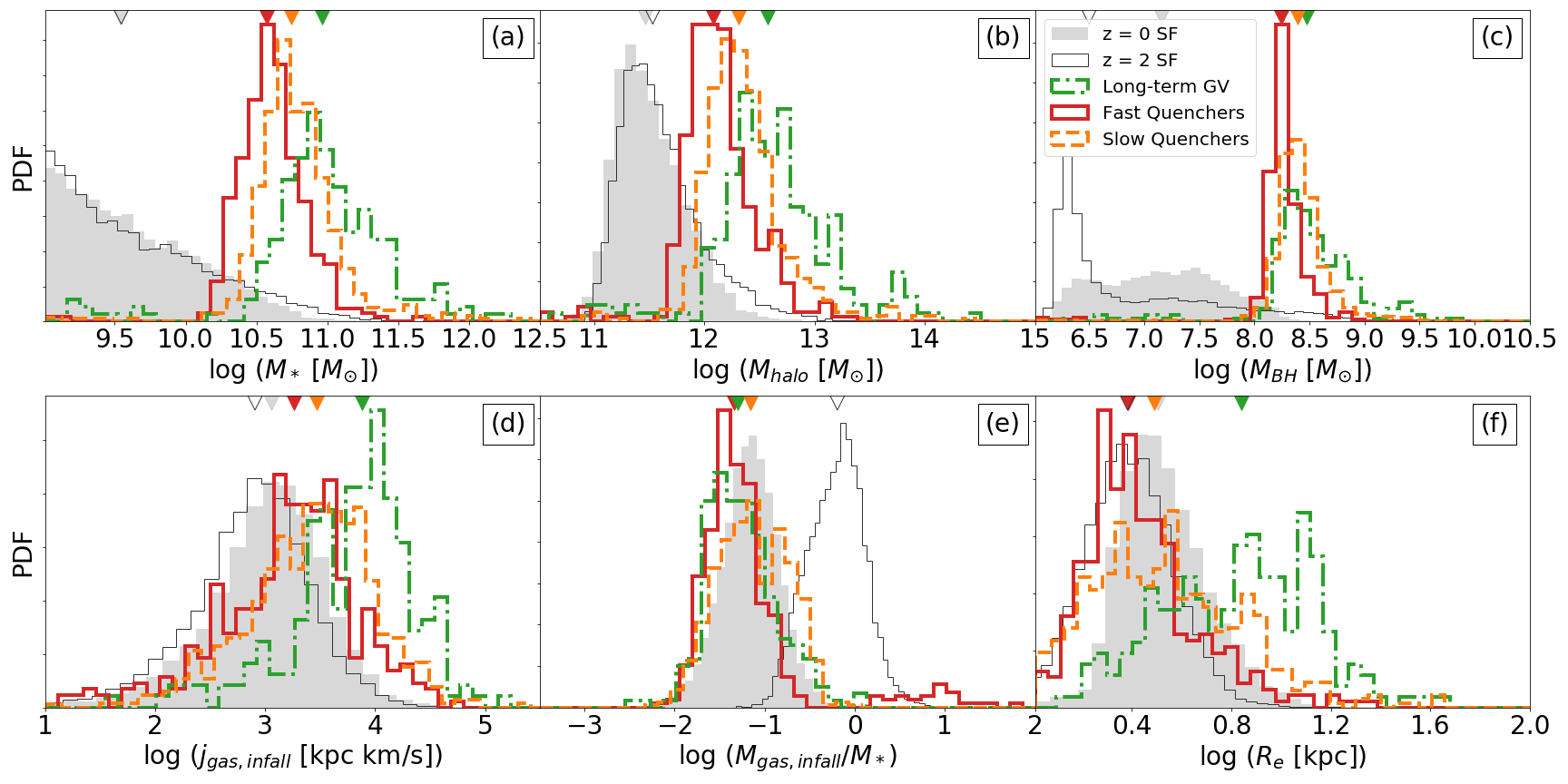}
\caption{\small Six properties of GV/Q subhaloes at their last snapshot prior to leaving the SFMS, which showed correlation with $\tq$ and $\tt$ in \Fig{lsf_cont_all}. Subhaloes are split into fast quenchers ($\tq < 1$ Gyr, red), slow quenchers ($\tq > 1$ Gyr, orange dashed), and long-term GV residents (or equivalently, very slow quenchers; $\tt > 3$ Gyr, green dash dotted). {{Additionally, SF galaxies are shown for reference at $z=0$ (grey shaded) and at $z=2$ (grey outline). Median values are indicated by triangles along the top.}} Panels starting at the top left: Stellar mass, total mass (including gas, stars, BHs, and DM) of the entire FOF halo, black hole mass, specific angular momentum of all gas particles infalling through $R_{\rm vir}/2$ in 100 Myr, total mass of the same gas particles, and stellar half-mass radius. In most properties (except infalling gas mass), the population of slow quenchers lies between fast quenchers and long-term GV residents, although there is significant overlap between all three groups. Subhaloes that quench faster generally have lower $\Ms$ and $M_{\rm halo}$. Most subhaloes leave the SFMS when $\MBH$ reaches $\sim10^{8.2}\Msun$, although slower (more massive) subhaloes are able to grow their BHs to modestly higher values prior to leaving the SFMS. The slower quenchers live in deeper potentials and begin quenching with more massive BHs. Fast quenchers tend to accrete relatively lower mass in gas, and that gas has lower specific angular momentum than slow quenchers. There is a small population of fast quenchers with very high infalling gas mass. This suggests that wet minor mergers may precede some fast quenching events. Fast quenchers generally have smaller stellar $R_e$.}
\label{lsf_props}
\end{figure*}

Panel (f) of \Fig{lsf_cont_all} shows subhalo half-mass radius prior to quenching. Larger galaxies tend to quench slowly compared to smaller galaxies. This is consistent with the the relation between $\tq$ and the specific angular momentum of their accreting gas.  Gas with lower angular momentum will tend to funnel toward the centre of the galaxy before forming stars, decreasing the effective radius.  Panels (d) and (f) show that such galaxies tend to be fast quenchers. When gas accretes with higher angular momentum, we expect that more star formation will occur in a galaxy's outskirts, increasing $R_e$. This is broadly consistent with the results of \cite{Gupta2020}, who found that, in TNG, massive extended galaxies selected at $z=2$ tend to quench later and more slowly (see their Figure 3, right hand panel). We also note the reversal of this trend at small $R_e$ for $\tq$ only. As we will show shortly in \Fig{lsf_props}, this regime represents only a very small portion of our sample so we refrain from drawing inferences from this result.

The remainder of the properties show very little correlation with $\tq$ or $\tt$. To calculate the disc fraction, we first compute the total angular momentum vector of the subhalo. For each gas particle within $2R_e$, we then compute the component of the angular momentum parallel to the total subhalo angular momentum ($J_z$). Next we compute the angular momentum of the same gas particle assuming a circular orbit with the same energy ($J_c$). Particles are identified as part of the disc component when $J_z/J_c > 0.7$. The disc fraction is then the mass of all disc component gas particles divided by the total gas mass. We also tried other common definitions for the disc fraction and found similar results. We note the moderate negative value of the Spearman coefficient for the gas fraction inside 1 kpc. This appears to be driven by a significant population of subhaloes with zero gas inside these radii, with a corresponding wide range of $\tq$ and $\tt$ values. Since the running medians themselves are fairly flat, we assess that this correlation is not significant.

To complement the above analysis of \Fig{lsf_cont_all}, we now split our sample of subhaloes into three groups. We define fast quenchers as those with $\tq <$ 1 Gyr (386 total in our sample), slow quenchers as those with $\tq >$ 1 Gyr (569 total), and long-term GV residents as those with $\tt >$ 3 Gyr (213 total). We note that our definitions exclude some subhaloes that have left the SFMS recently but have not yet quenched (those with $\tt <$ 3 Gyr), since their status as fast or slow quenchers is yet to be determined. Long-term GV residents may eventually become slow quenchers (but not fast quenchers). We classify them separately because their exact quenching duration is undetermined and $\tt$ is essentially a lower limit on their future $\tq$. We tried varying these definitions and found no significant change in our results.

\Fig{lsf_props} shows histograms of six properties, which showed significant correlation with $\tq$ and $\tt$ in \Fig{lsf_cont_all} (Spearman $\rho >$ 0.2 or clear visual correlation in the case of $R_e$), separated into our three classes of quenching duration. The histograms more clearly illustrate the parameter space in which most galaxies begin quenching, and also show the relationship between long-term GV residents and quenched galaxies in a straightforward way. We note that there is a fair amount of overlap between the histograms, and likewise scatter in the continuous plots, in all cases. However in most properties (except the infalling gas mass in panel e), the distribution for slow quenchers lies between the fast quenchers and the long-term GV residents. 

{{We also include in all panels of \Fig{lsf_props} the SF population at $z=2$ (grey outline) and $z=0$ (grey shaded) for reference. The three quenched populations lie roughly where we expect compared to SF galaxies: they have higher mass, higher mass BHs, and less gas infalling with more angular momentum. {There is significant evolution of the SF $\MBH$ distribution between $z=2$ and $z-0$, but at the same time there is very little evolution of the SF $\Ms$ distribution. While concerning, we will show that the same distributions for GV and Q galaxies do not evolve much. In other words, the TNG model may produce too aggressive evolution over cosmic time of the $\MBH$-$\Ms$ relation at low mass, but closer to the quenching mass regime the results are reasonable.} The infalling gas mass for SF galaxies decreases substantially over time, in agreement with the decline of the cosmic accretion rate over time (see, for example, \citealp{Chen2020}). $R_e$ increases modestly in agreement with the size evolution of galaxies (e.g., \citealp{VanDerWel2014}). }}

The first two panels of \Fig{lsf_props} are histograms of the stellar mass and the halo mass for our three classes of quenching duration. \fig{lsf_cont_all} showed that less massive galaxies in less massive halos tend to quench more quickly; \fig{lsf_props} shows the complementary relationship: fast quenchers tend to be less massive and live in less massive halos. The histograms also show that long-term GV residents tend to be the most massive galaxies. 

Panel (c) of \Fig{lsf_props} shows histograms of BH mass.  As noted earlier, virtually all quiescent and long-term GV galaxies have BH masses higher than $\MBH \sim10^{8.2}\Msun$ since BHs tend to switch to kinetic feedback at this mass. Slow quenchers tend to have slightly higher BH masses prior to quenching, which is consistent with their higher $\Ms$. We note that the histograms paint a slightly more nuanced picture than the same panel in \Fig{lsf_cont_all}. Although $\tq$ appears to be a continuous function of $\MBH$ in \Fig{lsf_cont_all}, the histogram in \Fig{lsf_props} shows that (at least for fast quenchers), the majority of galaxies leave the SFMS at essentially the same $\MBH$ as shown by the narrow peak in the histograms. Slow quenchers and long-term GV subahloes are able to build their $\MBH$ modestly higher (but in most cases $<$ 0.5 dex) prior to leaving the SFMS. 

Taking the first three panels  of \twofigs{lsf_cont_all}{lsf_props} together, it seems that in TNG, slower quenchers, which tend to be more massive, require more massive BHs in order to begin quenching.  Their deeper potentials are able to hold on to their gas longer following the onset of kinetic BH feedback at $10^8 \Msun$. {We have confirmed that $\tq$ and $\tt$ do indeed increase with increasing escape velocity at $R_e$ (similar to the first two panels of \Fig{lsf_cont_all}, not shown for brevity).} They are therefore able to continue to grow both their BHs and their stellar mass prior to, eventually, quenching.

Panels (d) and (e) of \fig{lsf_props} show the complementary picture for accreting gas as shown in the same panels of \fig{lsf_cont_all}.   Fast quenchers tend to accrete less gas than slow quenchers, and that gas tends to have lower specific angular momentum. We also note with interest that there is a small population of fast quenchers with high infalling gas mass. We do not see this population at $z=0$. Since fast quenchers also tend to accrete gas with low specific angular momentum, this suggests that some fast quenching events may be preceded by wet minor mergers. However, these galaxies contribute insignificantly to the total population of fast quenchers.  

{Angular momentum is expected to naturally increase with mass \citep{Peng2020,Renzini2020}, so it is not unexpected that slower quenchers, with higher mass, should have gas accreting with higher $j$.  To check whether the angular momentum of accretion correlates with $\tq$ independently of $\Ms$, we  calculated $\Delta j$, the difference between $j$ and the median $j$ for a given mass at a given redshift. We find that $\Delta j$ at $t_{\rm LSF}$ is weakly correlated with $\tt$ or $\tq$. However, by $z=0$, $j$ for all quenched galaxies is consistent with their expected distributions given their mass.}

The last panel of \Fig{lsf_props} shows histograms of half-mass radius for our three classes of quenching duration. In \fig{lsf_cont_all} we see that larger galaxies tend to quench more slowly than smaller galaxies, while here we find the complementary relationship: slow quenchers tend to be larger than fast quenchers. Long-term GV residents tend to be the largest galaxies. {Similar to our test for angular momentum, we also computed $\Delta R_e$ to check whether the variation in $R_e$ was due to the $R_e-\Ms$ scaling relation. In this case, we confirmed that fast quenchers were smaller than similar mass SF galaxies prior to quenching.} {We note with interest that $R_e$ (and $\Delta R_e$) are better predictors of $\tq$ than $j$ (and $\Delta j$).}

In summary, we find that in TNG, prior to leaving the SFMS, fast quenching galaxies tend to have (compared to slow quenching galaxies): lower stellar mass, less massive halos, less massive BHs, less infalling gas (although a few have very high infalling gas mass), less specific angular momentum in their infalling gas, and smaller effective radii. Galaxies that have lived in the GV for very long times occupy the opposite end of this spectrum of properties.

\section{Subhalo Properties at \MakeLowercase{z} = 0}
\label{secz0}

We have seen that most galaxies in TNG take $\ltsima 2$ Gyr to quench, but a significant number stay in the GV for much longer, some never fully quenching.  The slow quenchers leave the SFMS with higher mass, higher BH mass, larger sizes and higher angular momentum accreting gas compared to the faster quenchers.  Having established the properties of quenching galaxies prior to leaving the SFMS, do the properties of any of our groups evolve significantly by $z=0$? Put another way, are the $z=0$ properties of quenched galaxies indicative of their past quenching timescales?

\begin{figure*}
\includegraphics[width=1.0\linewidth]{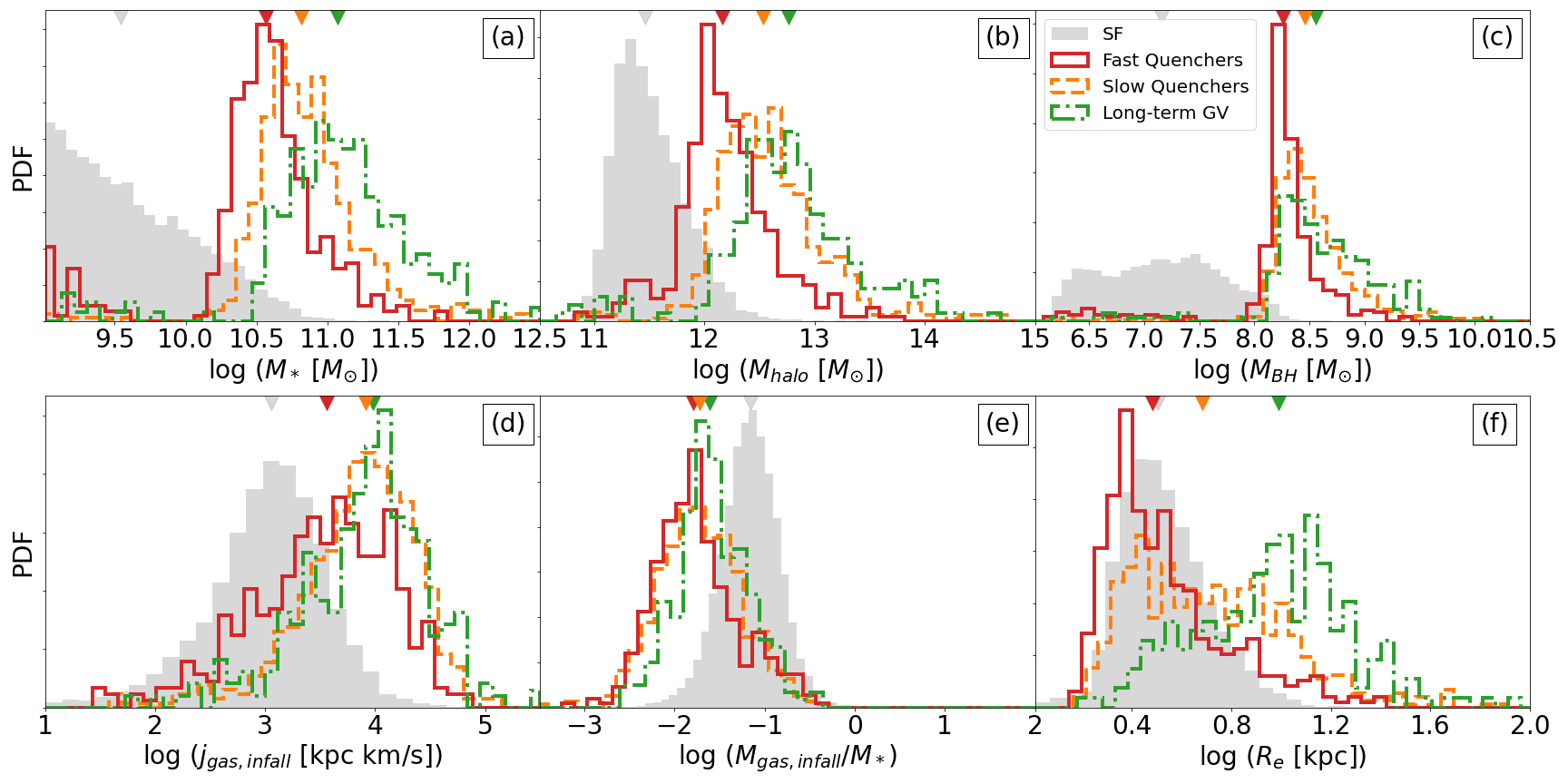}
\caption{\small Six properties of GV/Q subhaloes at $z=0$. Subhaloes are split into fast quenchers (red), slow quenchers (orange dashed), and GV residents (or equivalently, very slow quenchers; green dash dotted). {{Additionally, SF galaxies are shown for reference at $z=0$ (grey shaded). Median values are indicated by triangles along the top.}} Panels starting at the top left: Stellar mass, total mass (including gas, stars, BHs, and DM) of the entire FOF halo, black hole mass, specific angular momentum of all gas particles infalling through $R_{\rm vir}/2$ in 100 Myr, total mass of the same gas particles, and stellar half-mass radius. Comparing to \Fig{lsf_props}, there is very little evolution for most properties. The exception is the angular momentum of infalling gas (panel d), which is more comparable for all three groups.}
\label{z0_props}
\end{figure*}

\Fig{z0_props} shows the same properties of the same subhaloes as \Fig{lsf_props}, but at $z=0$. {{The first three}} properties are largely unchanged, suggesting TNG subhaloes evolve very little after quenching. {{The median values of $\Ms$, $M_{\rm halo}$, and $\MBH$ for the populations (shown at the top of \twofigs{lsf_props}{z0_props}) evolve by less than 2\%. Moreover, the change in $\Ms$ for individual galaxies between leaving the SFMS and either quenching or reaching $z=0$ was less than 1\% for all quenched galaxies and long-term GVs with $\tt < 3$ Gyr. There was a detectable increase in the mass of long-term GVs with $\tt>3$ Gyr, but the median increase was modest (only about 17\%.)}}.

{{Three properties did change, on average, between \twofigs{lsf_props}{z0_props}, indicating some redshift evolution. The first is}} the specific angular momentum of infalling gas (panel d). At $z=0$, all three groups show similar distributions that are weighted toward high specific angular momentum. Compared to \fig{lsf_props} the red and orange histograms  have moved up to coincide with the green histogram{{, and all lie higher than the SF population}}.  In other words, the fast and slow quenchers, which were once accreting gas with lower specific angular momentum, are now accreting gas with higher specific angular momentum just as the long-term GV galaxies. Other simulation studies have shown that the angular momentum of infalling streams increases with time \citep{Danovich2015,Stewart2017}, and thus the long-term GV may have been accreting gas with unusually high angular momentum prior to quenching.

{{The second property that shows redshift dependence is the infalling gas mass. Prior to quenching, all three groups were located about 1 dex below the $z=2$ SF population. At $z=0$, they again sit about 1 dex below the corresponding SF population. So although we see a significant evolution in the infalling gas mass for the quiescent galaxies, this change mirrors the same decrease in infalling gas to the SF galaxies. Therefore this evolution is roughly consistent with the overall change in available gas over cosmic time.}}

{The third property of quiescent galaxies that evolved over cosmic time is $R_e$. Again, the increase in median $R_e$ for our quenched groups is roughly consistent with the increase in $R_e$ of SF galaxies from $z=2$ to $z=0$, and with the observed size evolution of galaxies (e.g., \citealp{VanDerWel2014}). As with our analysis prior to quenching, $R_e$ values at $z=0$ for faster quenchers are lower than typical SF galaxies at the same mass.}

Regardless, TNG makes several directly testable predictions. At low redshift, fast quenchers in the TNG model end up with lower $\Ms$, lower $\MBH$, and smaller $R_e$. We will return to these predictions in the next section.

\begin{figure}
\includegraphics[width=1.0\linewidth]{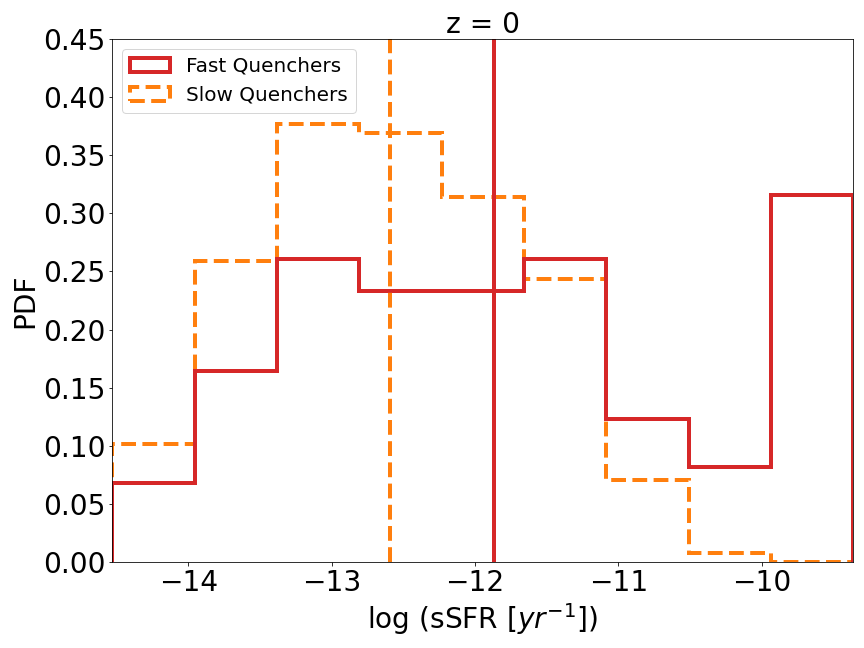}
\caption{\small sSFR at $z=0$ of all subhaloes that have ever quenched and still have non-zero sSFR, split into fast quenchers ($\tq <$ 1 Gyr, red) and slow quenchers ($\tq >$ 1 Gyr, orange dashed). Vertical lines indicate median values. A distinct population of rejuvenated fast quenchers lies at high sSFR, driving the median sSFR up for that group. However, the distributions of subhaloes that remain quenched (i.e. low sSFR) are fairly similar for both groups. Furthermore, more fast quenchers have zero (unresolved) sSFR than slow quenchers.}
\label{z0_ssfr}
\end{figure}

There are three other properties that differ between our three groups at $z=0$ that did not differ (much) prior to quenching: sSFR, gas fraction, and gas disciness. First, we examine the sSFR of the two quenched groups (i.e. those that have fully quenched at least once) that still have some residual sSFR today. \Fig{z0_ssfr} shows the residual sSFR at $z=0$ for all subhaloes that have ever quenched, split into fast and slow quenchers. There is a distinct population of fast quenchers which have rejuvenated, having fairly high sSFR today (i.e. on the SFMS). Rejuvenation is more common in fast quenchers. About $13^{+4}_{-3}$\% (Poisson errors) of fast quenchers will experience rejuvenation, compared with only $4^{+2}_{-2}$\% of slow quenchers. We defer investigation of any link between $\tq$ and rejuvenation rate to future studies. We speculate that perhaps low mass BHs can be powerful enough to temporarily expel central star-forming gas from low-mass (i.e. shallow potential) galaxies, but for some reason are less effective at heating the CGM and preventing gas infall than more massive BHs.  {{There is also the possibility that fast quenchers simply have more time to experience rejuvenation, since they spend less cosmic time in the quenching process.}} The rejuvenated galaxies push the median sSFR for fast quenchers slightly higher than the slow quenchers. 
However only 27\% of fast quenchers have non-zero sSFR compared to 43\% of slow quenchers.  Furthermore, in all parameters shown in \twofigs{lsf_props}{z0_props}, the slow quenching population lies between the fast quenchers and the long-term GV residents, which by definition have higher residual sSFR. In other words, the long-term GV residents are essentially very slow quenchers. Therefore we conclude that in TNG, fast quenchers are more likely to quench completely compared to slow quenchers, although they are also more likely to rejuvenate. 

\begin{figure}
\includegraphics[width=1.0\linewidth]{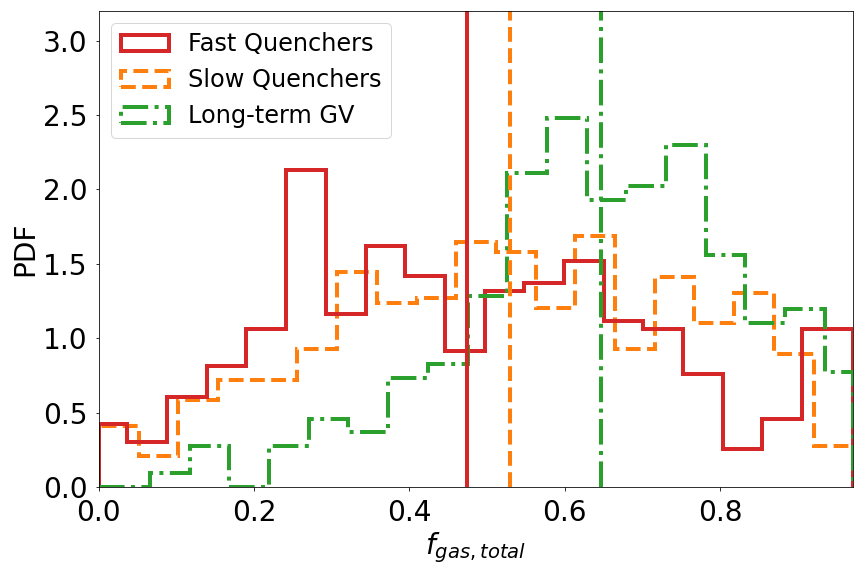}
\caption{\small Total bound gas fraction for our three GV/Q groups at $z=0$. Note that values are higher than typically seen for $f_{\rm gas}$ within 2$R_e$, as we are including all SUBFIND bound gas and stars. Vertical lines indicate median values. Fast quenchers tend to have less bound gas at $z=0$.}
\label{tng_gf}
\end{figure}

We next examine the gas fraction of our GV/Q subhaloes at $z=0$. \Fig{tng_gf} shows the total bound (i.e. SUBFIND) gas fraction for our three groups at $z=0$. Slower quenchers tend to have more gas, which {{could contribute}} to their longer quenching times. {{On the other hand, it could be that slower quenchers have more gas simply because they are consuming it more slowly.}}

\begin{figure}
\includegraphics[width=1.0\linewidth]{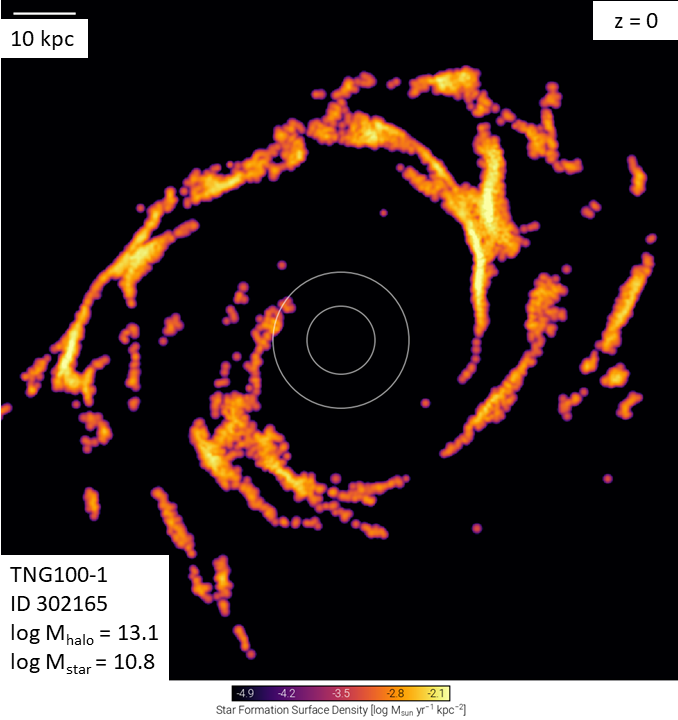}
\caption{\small SFR surface density for a typical long-term GV resident at $z=0$ (TNG100-1 Subfind ID 302165, log $\Ms$ = 10.8, $R_e$ = 5.7 kpc). White circles indicate $R_e$ and 2$R_e$. Scale is in the top left. Most GV/Q subhaloes have negligible gas content inside 2$R_e$, and the residual star formation that does occur tends to be in an extended ring. This image was created using the visualization tool available on the TNG Public Data Access website.}
\label{gf_example}
\end{figure}

Our choice to use total bound gas fraction is important. We find that most of our subhaloes have negligible gas fractions inside 1-2 kpc or even inside 1-2 $R_e$. Indeed, visual examinations of our long-term GV residents show that any residual star formation usually occurs outside of 2$R_e$ (see \Fig{gf_example} for an example, where the white circles indicate $R_e$ and $2R_e$).  Such residual star formation may be difficult to detect, especially in galaxies at higher $z$, although it is not unexpected in GV galaxies (see e.g. \citealp{Salim2012,Dekel2020}).

\begin{figure}
\includegraphics[width=1.0\linewidth]{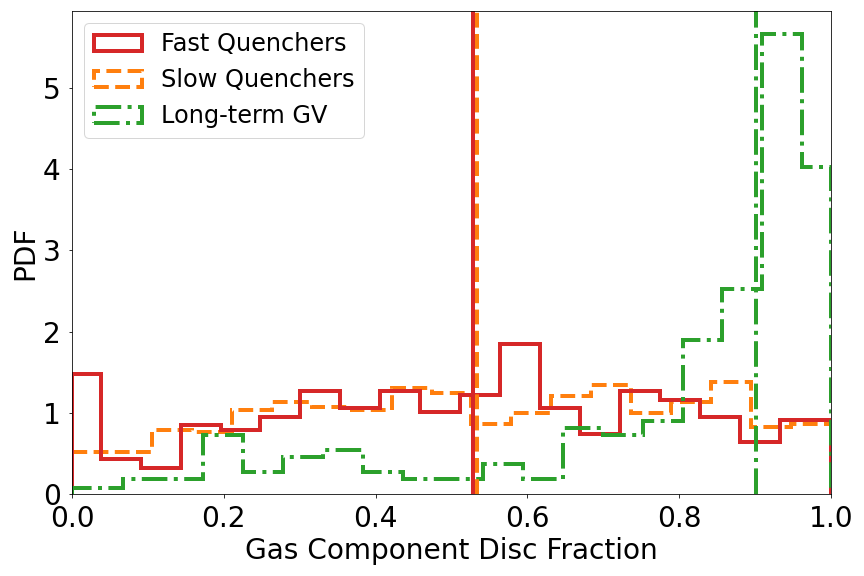}
\caption{\small Disciness of the gas component inside $r<2R_e$ for our three groups at $z=0$. Vertical lines indicate median values. GV residents tend to have very discy gas components.}
\label{disciness}
\end{figure}

Lastly, we examine the shape of our subhaloes in terms of disc fraction. \Fig{disciness} shows the disc fraction of the gas component for our three groups. The stellar components for all three groups are fairly similar (not shown). However, the disc fraction of the gas component within $2R_e$ for the long-term GV residents tends to be higher than in the quenched groups.  Fast and slow quenchers have similar distributions of disc fractions.

In summary, the distributions of most properties for fast and slow quenchers and the long-term GV residents remain largely unchanged by $z=0$.  The exception is the infalling gas angular momentum, which increases for fast and slow quenchers.  By $z=0$, fast quenchers are more likely to have experienced a rejuvenation event, but generally have lower gas fractions.  Long-term GV residents have higher disk gas fractions and are weakly star-forming in large extended disks.

\section{Discussion}
\label{secdiscussion}

\subsection{Comparison to Observations}
\label{secmanga}
 
\begin{figure}
\includegraphics[width=1.0\linewidth]{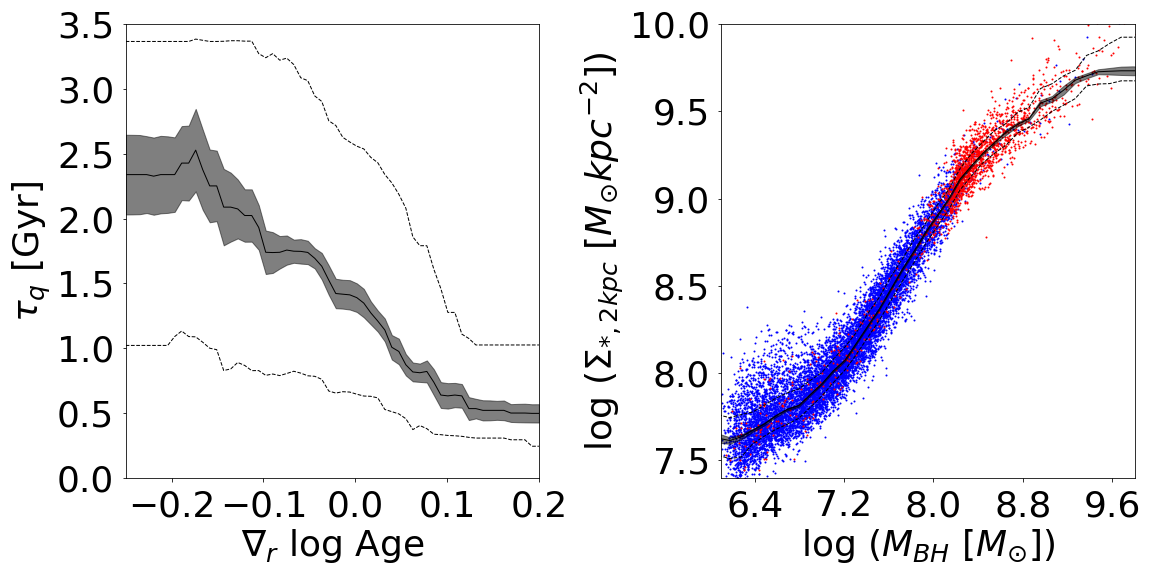}
\caption{\small First panel: Quenching duration $\tq$ vs stellar age gradient for TNG100-1 centrals. Since TNG subhaloes quench inside-out, slow quenchers end up with steeper (more negative) age gradients. Although the correlation has some scatter, the age gradient can be used as a crude substitute for $\tq$. Second panel: $\sigtwo$ vs $\MBH$ for TNG100-1 centrals with $\Ms > 10^9 \Msun$. The solid line is the running median value, shaded areas are the error on the median, and dashed lines are quartiles. Blue dots indicate SF galaxies while red dots indicate GV/Q galaxies. $\MBH$ and $\sigtwo$ are tightly correlated in TNG. }
\label{var_changes}
\end{figure}

Perhaps the most surprising thing we learned about TNG subhaloes is that the slow quenchers are more massive than fast quenchers, both before leaving the SFMS as well as at $z=0$.  Here we argue that our results are consistent with what is observed in the real universe at $z=0$.  In order to compare our results to observations, we must translate the duration of quenching to a measureable quantity.  Since most galaxies seem to quench inside-out \citep{Tacchella2015a,Tacchella2016,Ellison2018,Bluck2020,Nelson2021}, faster quenching should result in flatter stellar age gradients while slower quenching should result in steeper negative age gradients.  
{{For example, broadly speaking, a very fast inside-out quenching event would cease star formation at all radii at close to the same time, leaving behind similar stellar populations throughout the galaxy. A slow inside-out quenching process would leave older stars in the centre of the galaxy while younger stars are still forming in the outskirts, so once all star formation has ceased the imprint of a negative stellar age gradient should remain.}}
In \Fig{var_changes} (left) we show the relation between $\tq$ and stellar age gradient for TNG and confirm that they are negatively correlated (see also \citealp{Suess2021}). Therefore, we use the stellar age radial gradient as a proxy for $\tq$.  Second, we use $\sigtwo$ (the stellar surface density inside 2 kpc) as a proxy for $\MBH$ since they are expected to be tightly correlated \citep{Chen2020}, as they are in TNG (\Fig{var_changes}, right).  (However note that BH masses for SF galaxies are too massive in TNG by almost 2 orders of magnitude, although they agree reasonably well for Q galaxies - see Fig. 7 of \citealp{Terrazas2020}). 

We now compare our TNG results to observational data from \cite{Woo2019} who measured stellar age gradients as well as central stellar surface densities for galaxies in the MaNGA IFU survey.
Our sample consists of 263 quiescent central galaxies with 9.4 $<$ log $\Ms/\Msun$ $<$ 11.5. Stellar masses are taken from \cite{Brinchmann2004}. The half-mass radii ($R_{\rm e,mass}$) and central stellar surface densities (within 2 kpc) are taken from \cite{Woo2019} who computed them from the stellar mass profiles.  sSFR was computed from the emission lines and stellar densities within binned spaxels out to 2$R_{\rm e,mass}$.

\begin{figure*}
\includegraphics[width=1.0\linewidth]{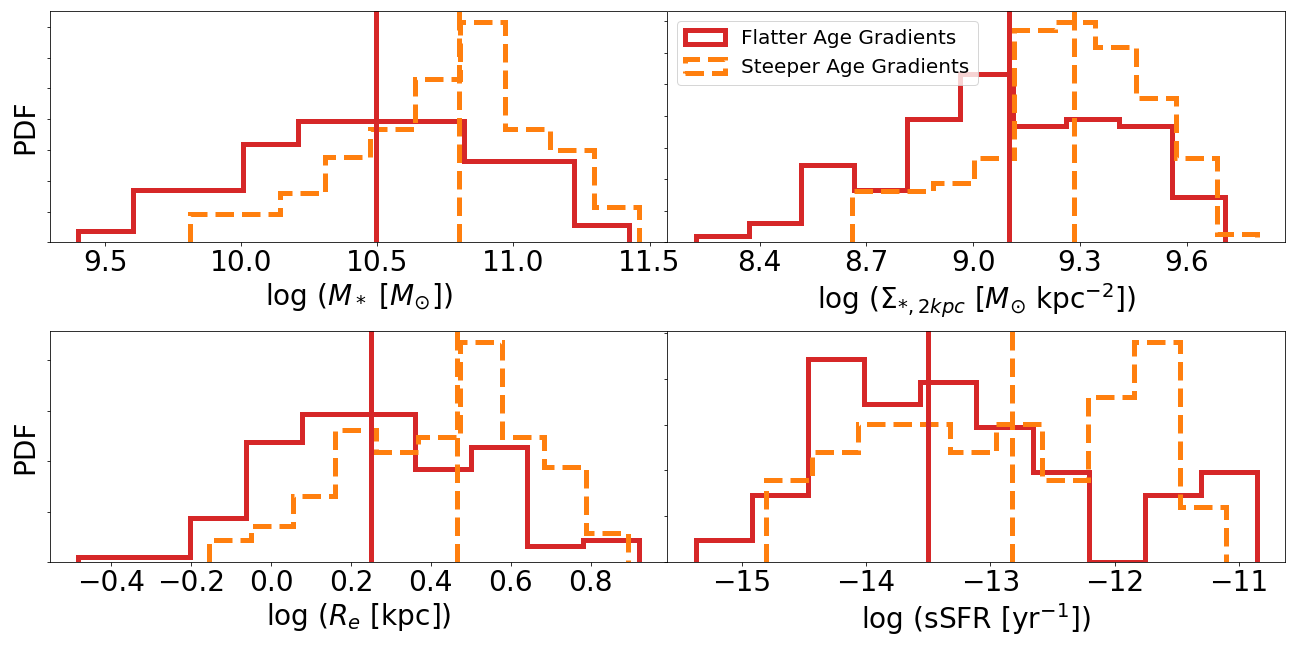}
\caption{\small Four properties of our sample of MaNGA central galaxies, split in half by stellar age gradient. Assuming inside-out quenching, flatter age gradients (red) should correspond to faster quenchers while steeper (i.e. more negative; orange dashed) age gradients should correspond to slower quenchers. Vertical lines indicate median values. All four properties agree with the predictions of TNG. Flatter age gradients have lower $\Ms$, lower $\sigtwo$ (our proxy for $\MBH$), smaller $R_e$, and lower detectable sSFR.}
\label{mangavars}
\end{figure*}

We split our sample of quiescent central MaNGA galaxies in half based on stellar age gradient and plot histograms of their $\Ms$, $\sigtwo$, $R_e$ and sSFR in \Fig{mangavars}.  Galaxies with flatter age gradients (i.e. faster quenchers) are shown in red and those with steeper age gradients (i.e. slower quenchers) are shown in orange. Due to our small sample size (263 galaxies - details of the sample are given in \citealp{Woo2019}), we do not split the galaxies into more groups nor do we attempt quantify the associated $\tq$. As with TNG, we see significant overlap between the populations. However, the qualitative trends agree. Central galaxies with steeper negative age gradients (i.e. longer $\tq$) generally have higher $\Ms$, higher $\sigtwo$ (i.e. $\MBH$), larger $R_e$, and higher residual sSFR (for those galaxies with any SF detected).

\begin{figure}
\includegraphics[width=1.0\linewidth]{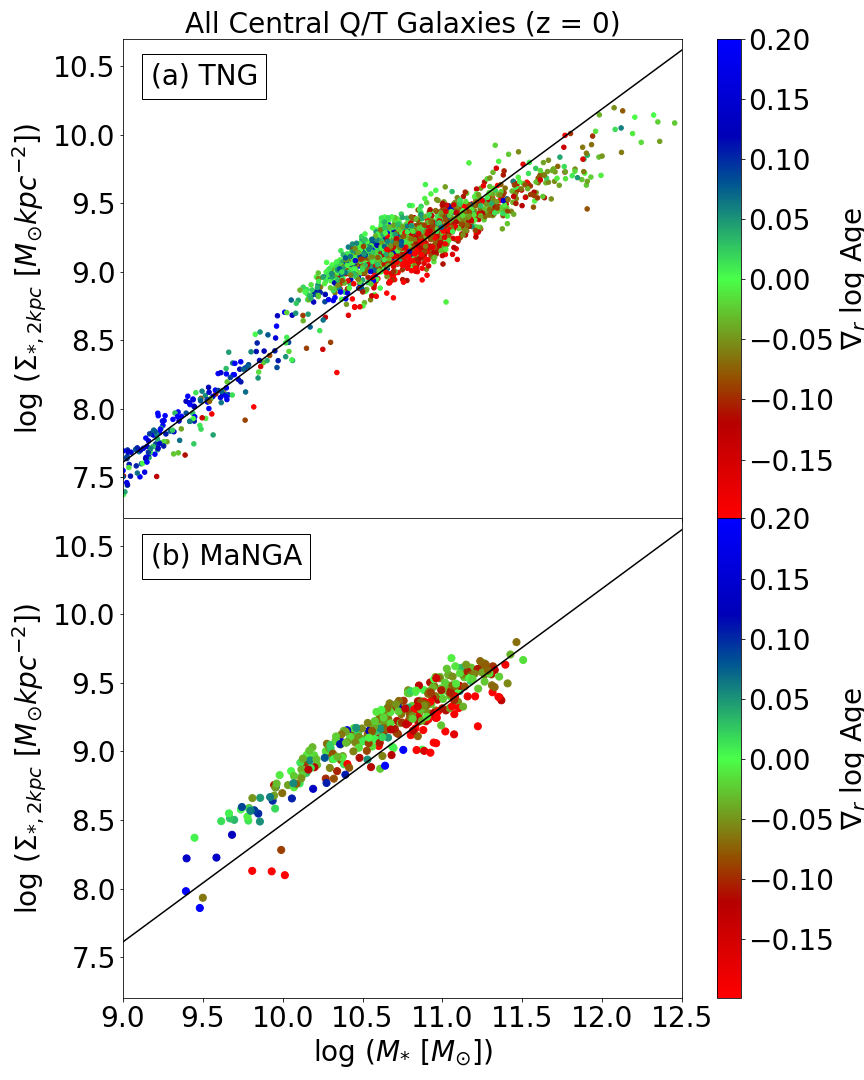}
\caption{\small Distribution of GV and Q galaxies in the $\sigtwo$-$\Ms$ plane for (a) TNG and (b) MaNGA. Points are color coded by the radial gradient of stellar age, with red points corresponding to steeper negative age gradients (i.e. slower quenchers). The line shown, for visual reference between both plots, is a least squares fit to the SF galaxies in TNG (not plotted). The agreement suggests that TNG's predictions are correct: slower quenchers are more massive and have higher $\sigtwo$ (a proxy for $\MBH$).}
\label{tngmanga}
\end{figure}

To emphasize the agreement between TNG and MaNGA, we show the age gradient of quiescent galaxies for both samples in the $\sigtwo$-$\Ms$ plane in \Fig{tngmanga}. Although the MaNGA sample is not mass-complete (nor, strictly, is the TNG sample), the agreement regarding the age gradients is striking. Both panels show that galaxies with steeper negative age gradients (i.e., the slower quenchers) are more massive and on average have higher $\sigtwo$.

Aside from our MaNGA comparison, several recent efforts to constrain quenching timescales using observations are worth noting. In particular, the work of 
\cite{Trussler2020} seems to corroborate our results.  These authors, improving on the work of \cite{Peng2015a}, studied observed stellar metallicities in local passive galaxies, and found that quenching is best explained by combination of gas starvation and outflows. In this scenario, low mass quenching is dominated by outflows and high mass quenching is dominated by starvation.  Of relevance to the current study, they found that quenching should take about 5-7 Gyr, with more massive galaxies generally quenching more slowly, in agreement with our results.  The comparison of our results with these studies is particularly apt since we have similarly taken a $z=0$ perspective of quenching timescale, i.e., examining the quenching timescales of galaxies that are quiescent today.

More recently, \cite{Tacchella2021} examined 161 observed galaxies at $z\sim0.8$, fitting both spectra and photometric data simultaneously in order to infer their star formation histories. They found that both the duration and epoch of quenching varies widely from galaxy to galaxy, with weak dependence on $\Ms$. 
Since they studied relatively high-redshift galaxies, a significant number of slow quenchers would likely not have been quiescent enough to enter their sample, and our results here predict that these will be of higher mass.  Furthermore, they are missing the low-$z$ fast quenchers which will mostly be of lower mass.  Hence their high-redshift study is not directly comparable to our $z=0$ study.

{\cite{Thomas2005,Thomas2010} measured the [$\alpha$/Fe] abundances in $z\sim 0$ early-type galaxies and found that the more massive ones had higher $\alpha$ abundances.  [$\alpha$/Fe] can be interpreted as a quenching timescale indicating that more massive galaxies had shorter quenching timescales.  However, higher abundances can also be interpreted as indicating shorter times on the SFMS rather than shorter quenching timescales so their results may not necessarily be in conflict with ours.  We defer the investigation of abundances to a future study.}

The results of \cite{Hahn2017} however are in direct conflict with ours.  These authors used a statistical model of quenching to match the observed bimodal sSFR distribution in local galaxies. Using Bayesian inference, and a set of priors on their free parameters, they found that galaxies should take $\sim$2-5 Gyr to quench, with more massive galaxies quenching more quickly, in contradiction to our findings.  Aside from the fact the sSFR distribution is likely not truly bimodal (which they acknowledge), in their model the mass dependence of $\tq$ (their parameter $A_\tau$) is strongly degenerate with another parameter that determines the mass dependence for the probability of quenching (their $A_{\rm P_Q}$).  The prior distribution of this parameter was restricted such that an anti-correlation between $\tq$ and mass was almost guaranteed.  It would be worth repeating their analysis with less restrictive priors to see if their model truly requires this anti-correlation.

\subsection{Comparison to Previous Simulation Studies}

As mentioned in the Introduction, several simulation papers have reported conflicting results regarding the dependence of the quenching timescale on epoch and mass \citep{Nelson2018a, Wright2018,Dave2019}.  Regarding the dependence on epoch, previous studies (both simulation and observational) defined the ``epoch" as either the epoch at which a galaxy finishes quenching, or the epoch at which the quiescent galaxy is observed (in which case, it is after it finished quenching).  Defined in this way, quenching timescales are shorter at earlier times simply because the age of the universe was lower at earlier times, which we confirm in the right panel of \Fig{tauq_vs_lookback}.  In this work, we have also shown the ``redshift zero" perspective by measuring the dependence of quenching timescale on the epoch at which quiescent galaxies (observed at $z=0$) began to quench.  Although there is a slight epoch dependence due to constraints of definition, we have found that the average quenching timescale {varies very little} over cosmic time.  The {weak dependance on epoch} of the quenching timescale reflects the fact that the primary quenching mechanism in TNG is kinetic black hole feedback, the conditions of which are not strongly dependent on cosmic time.  

Despite the constancy of the average quenching timescale, the scatter is dependent on several other parameters, most notably mass.  We have shown here that more massive quiescent galaxies took longer to quench than less massive quiescent galaxies.  It is here where other simulation papers seem to disagree.
The most notable apparent disagreement with our results is with \cite{Nelson2018a}, who also studied TNG100-1 but found shorter $\tq$ at high $\Ms$. We believe that the source of this discrepancy is their definition of the GV. They define the GV in terms of g-r colour by fitting the blue and red sequences with a mass- and redshift-dependent double Gaussian density profile. This results in a drastically narrower colour gap between blue and red loci above $\sim 10^{10} \Msun$ (see their Figure 4, bottom panel for $z=0$). In TNG, g-r colour is tightly correlated with sSFR at all masses. By our estimation, from their Figure 4 at $z=0$, a subhalo with $\Ms = 10^{10} \Msun$ would have a GV width of 1.8 dex in sSFR. A subhalo with $\Ms = 10^{11} \Msun$ would have a GV width of only 0.6 dex in sSFR. Although they mention the possibility that the shorter colour gap at high mass may contribute to their shorter $\tq$ at high mass, they primarily attribute it to higher $\MBH$ and therefore stronger feedback. Our analysis shows the opposite is true: subhaloes with more massive BHs in TNG actually quench more slowly as measured by sSFR.

Two other recent simulation papers are worth noting. First, \cite{Wright2018} studied quenching timescales in the EAGLE simulations (Evolution and Assembly of GaLaxies and their Environments) and found that more massive galaxies quench more quickly.  EAGLE is a set of cosmological hydrodynamical simulations with similar size, resolution, and sub-grid treatments as TNG \citep{McAlpine2016,Schaye2015,Crain2015}. EAGLE varies notably from TNG in that it uses a thermal (quasar mode) only recipe for AGN feedback \citep{Schaye2015}.  
\cite{Wright2018} used two definitions of the GV, one based on u-r colour and the other based on sSFR. We note that their u-r colour-based GV has a fairly uniform width across $\Ms$ (their Figure 3) and they obtain similar results using both definitions. They found maximum $\tq$ at $\Ms \sim 10^{9.7} \Msun$, with the shortest quenching durations at high $\Ms$. 

Second, \cite{RodriguezMontero2019} investigated quenching durations in the SIMBA simulations \citep{Dave2019}. SIMBA is similar to EAGLE and TNG in terms of size and resolution, but notably uses a bipolar kinetic feedback prescription for AGNs. They found quenching durations at most masses took $\sim 0.1 t_H$ (where $t_H$ is the age of the universe at quenching), except at $\Ms \sim 10^{10.3} \Msun$ where quenching was more rapid (roughly the mass at which AGNs become effective in SIMBA). In contrast to EAGLE and TNG, the quenching duration in SIMBA is strongly bimodal.

It appears that there could be a genuine tension between the EAGLE, SIMBA, and TNG models, which would be a tantalizing result if testable predictions can be extracted. However, the methods in these analyses are still significantly different from ours, making direct comparison difficult. In future work we intend to conduct an identical analysis on the EAGLE and SIMBA data to provide a fair comparison.

\subsection{Implications for Galaxy Evolution}

Our analysis of quenching timescales in TNG has revealed a surprising number of galaxies that quench slowly, some staying in the GV for many Gyrs.  Their properties immediately prior to quenching (Sec. \ref{seclsf}) give important clues to the factors that determine quenching duration.
Slower quenchers tend to be more massive and live in more massive halos than fast quenchers.  Although their BHs are more massive than the nominal threshold for kinetic feedback in TNG ($\sim10^8 \Msun$ - see for example \citealp{Terrazas2020,Zinger2020}), the star-forming gas lives in deeper potential wells, requiring longer times to {expel, keep out, and/or consume}.  Furthermore, slower quenchers accrete gas with higher angular momentum than fast quenchers.

\begin{figure}
\includegraphics[width=1.0\linewidth]{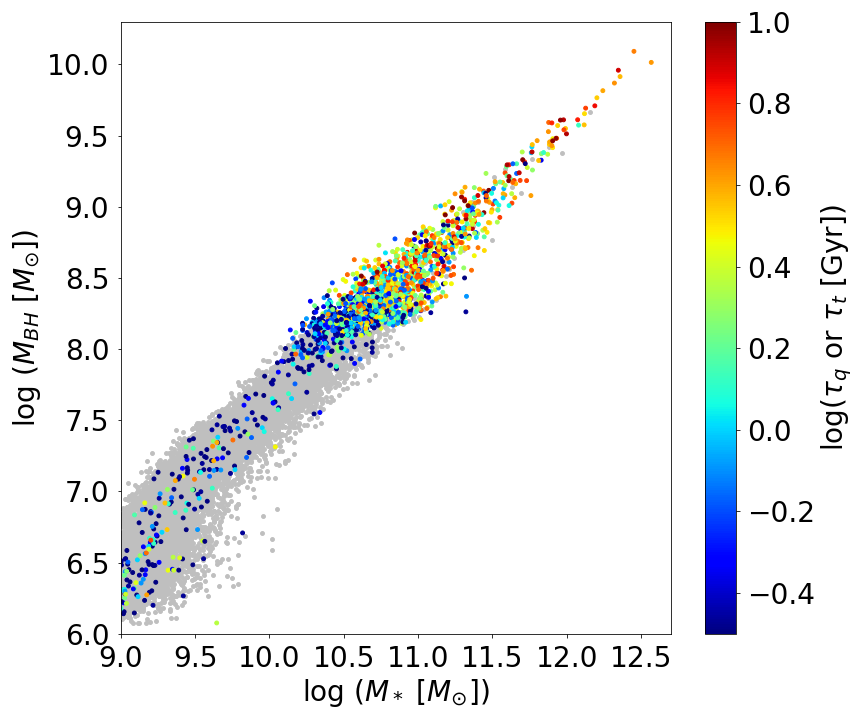}
\caption{\small All TNG centrals at $z=0$ in $\MBH$-$\Ms$. Grey dots are SF, while Q and GV galaxies are coloured by $\tq$ or $\tt$ respectively. TNG produces a quenching boundary (i.e. between grey and coloured points) that is mass-dependent, where more massive galaxies have more massive BHs in order to be quiescent.}
\label{mbh_mstar}
\end{figure}

The deeper potential wells of more massive galaxies also offers an explanation for the mass-dependence of the quenching boundary in the $\MBH$-$\Ms$ diagram.  \Fig{mbh_mstar} shows the TNG $\MBH$-$\Ms$ diagram at $z=0$.  Grey points are galaxies that are still star-forming.  The other points are quiescent and GV galaxies colour-coded by their quenching duration.  Although there are a handful of quiescent/transitioning galaxies (coloured points) below about $\MBH \sim 10^{8}\Msun$, the vast majority of them are concentrated above that BH mass, forming a relatively sharp boundary between star-forming and quiescent galaxies corresponding to the rough threshold between thermal and kinetic feedback.  Notice that this boundary is mass-dependent.  Although the mass-dependence is not as steep as observed (\citealp{Terrazas2016}), our results may hint at its origin: while still star-forming the slow quenchers (which are more massive) are accreting relatively more gas (\fig{lsf_props}) into deeper potential wells, which allows the BH to grow past $10^8 \Msun$ without quenching.  We have also shown that neither $\Ms$ nor $\MBH$ evolve significantly during quenching, and therefore, quiescent galaxies arrived near their current position in the $\MBH$-$\Ms$ diagram {\it while still star-forming}.

{{Although we have drawn particular attention here to potential well depth, we emphasize that the mass dependence of $\tq$ could very likely be driven by the combination of several factors. For a purely ejection driven scenario, we would expect quenching timescales on the order of $\sim$10 of Myr given typical outflow velocities in TNG \citep{Nelson2019a}. Gas depletion timescales, on the other hand, are on the order of less than 1 Gyr if we consider the gas within $R_e$, but much longer than the age of the universe if we consider all bound gas (even at $z=0$). Therefore it is unlikely that quenching in TNG is merely ejective, and moderated by potential depth.   Indeed, \cite{Zinger2020} showed that BH feedback in TNG can be both ejective and preventative, with feedback eventually heating the CGM to prevent infall. Furthermore, in the TNG model gas tends to be ejected only from the feedback region of 2-3 kpc from the centre, and it is ejected perpendicular to the plane of the galaxy and therefore does not directly push fuel out of the disc at high outflow rates \citep{Nelson2019a}.}}

{{An alternative explanation for the mass dependence of the quenching timescale may lie in the idea of angular momentum quenching (e.g. \citealp{Peng2020,Renzini2020}). In this scenario, more massive galaxies accrete gas with higher angular momentum, particularly later in cosmic time. This high angular momentum accretion could eventually lead to gas accumulating at large radii with high rotation, preventing its ability to cool, clump, and form stars. Even if angular momentum quenching proves not to be a primary quenching mechanism, it can explain the longer $\tq$ in more massive galaxies, which seem in TNG to experience rejuvenation less often.}}

\begin{figure}
\includegraphics[width=\linewidth]{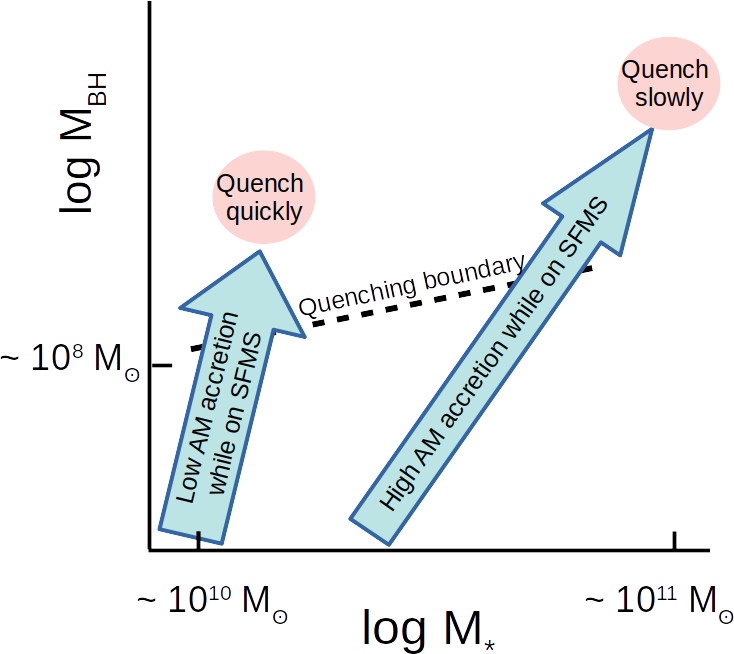}
\caption{\small Schematic diagram of evolutionary tracks in log $\MBH$-log $\Ms$ prior to quenching.  This schematic represents a portion of the $\MBH$-$\Ms$ diagram zoomed-in around the quenching boundary.  }
\label{cartoon}
\end{figure}

How do galaxies evolve in the $\MBH$-$\Ms$ diagram prior to quenching to bring them to their current positions?
Prior to quenching, the gas accreting onto fast quenchers tends to have lower angular momentum (\fig{lsf_props}d). As reported in \cite{Walters2021}, low angular momentum accretion leads to steeper growth in the $\sigtwo$-$\Ms$ plane (see their Fig. 18), or equivalently, the $\MBH$-$\Ms$ plane, since $\sigtwo$ and $\MBH$ are tightly correlated (\Fig{var_changes}, right).  {{\cite{Walters2021} showed that this effect is long-lived and is predictive of galaxy growth trajectory in $\sigtwo$-$\Ms$ at least up to 5 Gyr (from $z=0.5$ to today). Indeed, we see the signature of this growth in $j$, $R_e$, and in the gas disc fractions prior to quenching. Although all our quenching groups showed high disciness prior to quenching (consistent with SF galaxies), long-term GV residents had the highest median gas disc fractions, followed by slow quenchers and lastly fast quenchers (not shown for brevity).}} Therefore fast quenchers grow their $\MBH$ quickly relative to their total stellar mass and so end up with lower $\Ms$ when they first leave the SFMS.  In contrast, slow quenchers accrete gas with higher angular momentum, causing their stellar mass to grow faster than their $\MBH$, so they end up with higher $\Ms$ when they leave the SFMS.  We summarize this picture in \Fig{cartoon} which shows a schematic representation of the evolution of galaxies into quiescence on the $\MBH$-$\Ms$ diagram, zoomed-in around the quenching boundary.

In summary, it appears that in TNG, galaxy quenching duration is driven primarily by $\Ms$ and $\MBH$ at the onset of quenching, the combination of which are determined primarily by the angular momentum of accreting gas while star-forming.

\section{Summary}
\label{secsummary}

We investigated the quenching timescales of central galaxies in the IllustrisTNG simulations. Our results are summarized as follows:

\begin{enumerate}
 \item Roughly 40\% of the galaxies in TNG quench rapidly ($\tq < $ 1 Gyr). A significant tail of the $\tq$ distribution extends to longer times, up to $\sim$ 10 Gyr (\Fig{tauq_dist}, red line).
 \item TNG predicts another significant population of galaxies that leave the SFMS but do not completely quench within the age of the universe.  These galaxies reside in the GV for very long times until $z=0$ (\Fig{tauq_dist}, green line).
\item Quenching duration is slightly longer for galaxies that began to quench earlier, or that finished their quenching later, but overall the average quenching timescale {depends only weakly on} cosmic time (\Fig{tauq_vs_lookback}).
\item The absolute and fractional rates of galaxies leaving the SFMS increase steadily over cosmic time. Slow quenchers are about twice as common as fast quenchers at intermediate redshifts ($z \sim 2$ to $z \sim 0.7$, \Fig{epoch}).
\item Prior to quenching, galaxies that will quench slowly tend to have: higher $\Ms$, higher $M_{halo}$, higher absolute $\MBH$ and larger $R_e$ (\twofigs{lsf_cont_all}{lsf_props}). They accrete more gas, and with higher specific angular momentum.
\item We find no significant correlation between $\tq$ and sSFR, instantaneous BH accretion rate, or gas fraction prior to quenching (\Fig{lsf_cont_all}).

\item With the exception of the accreting gas angular momentum, the distributions of properties of quenched galaxies at $z=0$ are largely unchanged since leaving the SFMS, suggesting that quiescent galaxies in TNG evolve very little during and after quenching (\Fig{z0_props}).
\item At $z=0$, fast quenchers, which were accreting gas at lower angular momentum prior to quenching, are now acreting gas with the same high angular momentum as slow quenchers.
\item At $z=0$, galaxies that quenched slowly tend to have extended gas disks/rings with low star formation rates (Fig. \ref{disciness}).
\item A comparison with observations of nearby quiescent galaxies, using IFU derived age gradients as a proxy for $\tq$, shows broad agreement with TNG's predictions for $z=0$ $\Ms$, $\sigtwo$, $R_e$, and residual sSFR (\twofigs{mangavars}{tngmanga}).
\end{enumerate}

Our results imply that galaxies evolve as shown in the schematic diagram in \Fig{cartoon}. Prior to quenching, the angular momentum of accreting gas determines the evolutionary track of a galaxy in $\MBH$-$\Ms$ plane. In turn, this determines where a galaxy will reach the quiescent region of the diagram. Galaxies with deeper potentials take longer to quench, requiring more massive BHs to expel gas and prevent its infall.

Our comparison with observations suggests that TNG's model for AGN feedback and quenching is on the right track. We caution, however, that we have not yet performed the same analysis on other modern cosmological simulations. It is entirely possible that these results could be common to any model using AGN-driven gas ejection and infall prevention to quench galaxies.

\section*{Data Availability}

Data from the IllustrisTNG simulations are publicly available at \url{www.tng-project.org}. Data specific to this paper are available on request from the corresponding author.

\section*{Acknowledgements}

We sincerely thank the referee for the constructive feedback and excellent suggestions. We acknowledge the helpful and stimulating discussions with Salvatore Quai and Sandro Tacchella. SLE gratefully acknowledges the receipt of an NSERC Discovery Grant. 

The simulations of the IllustrisTNG project used in this work were undertaken with compute time awarded by the Gauss Centre for Supercomputing (GCS) under GCS Large-Scale Projects GCS-ILLU and GCS-DWAR on the GCS share of the supercomputer Hazel Hen at the High Performance Computing Center Stuttgart (HLRS), as well as on the machines of the Max Planck Computing and Data Facility (MPCDF) in Garching, Germany.

This research made use of Astropy,\footnote{http://www.astropy.org} a community-developed core Python package for Astronomy \citep{Astropy2013,Astropy2018}. 
This research also made use of the computation resources provided by Westgrid (www.westgrid.ca) and Compute Canada (www.computecanada.ca).  

Funding for the Sloan Digital Sky Survey IV has been provided by the Alfred P. Sloan Foundation, the U.S. Department of Energy Office of Science, and the Participating Institutions. SDSS acknowledges support and resources from the Center for High-Performance Computing at the University of Utah. The SDSS web site is www.sdss.org.  SDSS is managed by the Astrophysical Research Consortium for the Participating Institutions of the SDSS Collaboration including the Brazilian Participation Group, the Carnegie Institution for Science, Carnegie Mellon University, the Chilean Participation Group, the French Participation Group, Harvard-Smithsonian Center for Astrophysics, Instituto de Astrofísica de Canarias, The Johns Hopkins University, Kavli Institute for the Physics and Mathematics of the Universe (IPMU) / University of Tokyo, the Korean Participation Group, Lawrence Berkeley National Laboratory, Leibniz Institut für Astrophysik Potsdam (AIP), Max-Planck-Institut für Astronomie (MPIA Heidelberg), Max-Planck-Institut für Astrophysik (MPA Garching), Max-Planck-Institut für Extraterrestrische Physik (MPE), National Astronomical Observatories of China, New Mexico State University, New York University, University of Notre Dame, Observatório Nacional / MCTI, The Ohio State University, Pennsylvania State University, Shanghai Astronomical Observatory, United Kingdom Participation Group, Universidad Nacional Autónoma de México, University of Arizona, University of Colorado Boulder, University of Oxford, University of Portsmouth, University of Utah, University of Virginia, University of Washington, University of Wisconsin, Vanderbilt University, and Yale University.

\bibliographystyle{mnras}
\bibliography{library}

\label{lastpage}

\end{document}